\documentclass[apj]{emulateapj}
\newcommand{\etal}{{et al.~}}
\newcommand{\msunh}{\>h^{-1}\rm M_\odot}
\newcommand{\Lsunhh}{\,h^{-2}\rm L_\odot}
\newcommand{\Msun}{\>{\rm M_{\odot}}}

\newcommand{\mpch}{\>h^{-1}{\rm {Mpc}}}
\newcommand{\hmpc}{\>h{\rm {Mpc}}^{-1}}
\newcommand{\kms}{\>{\rm km}\,{\rm s}^{-1}}
\newcommand{\kmsmpc}{\>{\rm km}\,{\rm s}^{-1}\,{\rm Mpc}^{-1}}
\newcommand{\Qmag}{\>^{0.1}{\rm M}_Q-5\log h}
\newcommand{\rmag}{\>^{0.1}{\rm M}_r-5\log h}
\newcommand{\rrmag}{\>^{0.0}{\rm M}_r-5\log h}

\usepackage{xcolor}
\shorttitle{Lighting dark matter halos with galaxies}
\shortauthors{Yang et al.}

\begin{document}
            
%%%%%%%%%%%%%%%%%%%%%%%%%%%%%%%%%%%%%%%%%%%%%%%%%%%%%%%%%%%%%%%%%%%%%%%%%%

\title{ELUCID V. Lighting dark matter halos with galaxies }
    
\author{Xiaohu Yang\altaffilmark{1,2}, Youcai Zhang\altaffilmark{3},
  Huiyuan Wang\altaffilmark{4,5}, Chengze Liu\altaffilmark{1},
  Tianhuan Lu\altaffilmark{6}, Shijie Li\altaffilmark{1}, Feng
  Shi\altaffilmark{3}, Y.P. Jing\altaffilmark{1,2},
  H.J. Mo\altaffilmark{7,8}, Frank C. van den Bosch\altaffilmark{9},
  Xi Kang\altaffilmark{10}, Weiguang Cui\altaffilmark{11}, Hong
  Guo\altaffilmark{3}, Guoliang Li\altaffilmark{9},
  S.H. Lim\altaffilmark{7}, Yi Lu\altaffilmark{3}, Wentao
  Luo\altaffilmark{1}, Chengliang Wei\altaffilmark{9}, Lei
  Yang\altaffilmark{1} }

\altaffiltext{1}{Department of Astronomy, Shanghai Key Laboratory for
  Particle Physics and Cosmology, Shanghai Jiao Tong University,
  Shanghai 200240, China; E-mail: xyang@sjtu.edu.cn}

\altaffiltext{2}{IFSA Collaborative Innovation Center, and Tsung-Dao
  Lee Institute, Shanghai Jiao Tong University, Shanghai 200240,
  China}

\altaffiltext{3}{Shanghai Astronomical Observatory, Nandan Road 80,
  Shanghai 200030, China}

\altaffiltext{4}{Key Laboratory for Research in Galaxies and
  Cosmology, Department of Astronomy, University of Science and
  Technology of China, Hefei, Anhui 230026, China}

\altaffiltext{5}{School of Astronomy and Space Science, University of
  Science and Technology of China, Hefei 230026, China}

\altaffiltext{6}{Zhiyuan College, Shanghai Jiao Tong University,
  Shanghai 200240, China}

\altaffiltext{7}{Department of Astronomy, University of Massachusetts,
  Amherst MA 01003-9305, USA}

\altaffiltext{8}{Astronomy Department and Center for Astrophysics,
  Tsinghua University, Beijing 10084, China}

\altaffiltext{9} {Astronomy Department, Yale University, P.O. Box
  208101, New Haven, CT 06520-8101}

\altaffiltext{10} {Purple Mountain Observatory, the Partner Group of
  MPI für Astronomie, 2 West Beijing Road, Nanjing 210008, China}

\altaffiltext{11}{Departamento de F\'isica Te\'orica, M\'odulo 15,
  Facultad de Ciencias, Universidad Aut\'onoma de Madrid, E-28049
  Madrid, Spain}

%%%%%%%%%%%%%%%%%%%%%%%%%%%%%%%%%%%%%%%%%%%%%%%%%%%%%%%%%%%%%%%%%%%%%%%%%%

\begin{abstract}
  In a recent study, using the distribution of galaxies in the north
  galactic pole of SDSS DR7 region enclosed in a 500$\mpch$ box, we
  carried out our ELUCID simulation (Wang et al. 2016, ELUCID
  III). Here we {\it light} the dark matter halos and subhalos in the
  reconstructed region in the simulation with galaxies in the SDSS
  observations using a novel {\it neighborhood} abundance matching
  method.  Before we make use of thus established galaxy-subhalo
  connections in the ELUCID simulation to evaluate galaxy formation
  models, we set out to explore the reliability of such a link. For
  this purpose, we focus on the following a few aspects of galaxies:
  (1) the central-subhalo luminosity and mass relations; (2) the
  satellite fraction of galaxies; (3) the conditional luminosity
  function (CLF) and conditional stellar mass function (CSMF) of
  galaxies; and (4) the cross correlation functions between galaxies
  and the dark matter particles, most of which are measured separately
  for all, red and blue galaxy populations.  We find that our
  neighborhood abundance matching method accurately reproduces the
  central-subhalo relations, satellite fraction, the CLFs and CSMFs
  and the biases of galaxies. These features ensure that thus
  established galaxy-subhalo connections will be very useful in
  constraining galaxy formation processes. And we provide some
  suggestions on the three levels of using the galaxy-subhalo pairs
  for galaxy formation constraints. The galaxy-subhalo links and the
  subhalo merger trees in the SDSS DR7 region extracted from our
  ELUCID simulation are available upon request.
\end{abstract}

%%%%%%%%%%%%%%%%%%%%%%%%%%%%%%%%%%%%%%%%%%%%%%%%%%%%%%%%%%%%%%%%%%%%%%%%%%


%%%%%%%%%%%%%%%%%%%%%%%%%%%%%%%%%%%%%%%%%%%%%%%%%%%%%%%%%%%%%%%%%%%%%%%%%%

\section{Introduction}

To fully model the structure formation of the universe and to probe
the detailed galaxy formation processes, one needs to have a fair
sampling of the universe with sufficient large volume and resolution.
Thanks to large redshift surveys, such as the Sloan Digital Sky Survey
\citep[SDSS;][]{York_etal00}, we are now able to carry out such kind
of investigations to an unprecedented accuracy.  However, in order to
make full use of the potential of the observational data, one still
has to develop or make use of optimal strategies. One of the most
efficient ways of using the observational data is to carry out
constrained simulations where the initial density field is indeed
extracted from the observations, which is the basic idea of our ELUCID
project.

Along this line, numerous attempts have been made to develop methods
to reconstruct the initial conditions of structure formation in the
local universe using galaxy distributions and/or peculiar velocities
\citep{Sousa2007, Jasche2013, Jasche2015, Sousa2015, Seljak2017}.
\citet{HoffmanRibak91} developed a method to construct Gaussian random
fields that are subjected to various constraints \citep[see
also][]{Bertschinger87, vandeWeygaert96, Klypin_etal03,
  KitauraEnblin08}.  \cite{Klypin_etal03} improved this method by
using Wiener Filter \citep[see e.g.][]{Zaroubi_etal95} to deal with
sparse and noisy data.  Gaussian density fields constrained by the
peculiar velocities of galaxies in the local universe have also been
used to set up the initial conditions for constrained simulations
\citep[e.g.][]{Kravtsov_etal02, Klypin_etal03, Gottloeber_etal10}.
Note, however, that the basic underlying assumption in this method is
that the linear theory is valid on all scales \citep{Klypin_etal03,
  Doumler_etal13}.

In a recent paper, \citet[][hereafter ELUCID I]{Wang_etal14} developed
a method combining the Bayesian reconstruction approach with a much
more accurate dynamic model of structure evolution, the Particle Mesh
(PM) model. The PM technique has been commonly adopted in $N$-body
codes to evaluate gravitational forces on relatively large scales
\citep[see e.g.][] {White_etal83, KlypinShandarin83, JingSuto02,
  Springel2005b}, and can follow the structure evolution accurately as
long as the grid cells and time steps are chosen sufficiently small.
Tests show that this method can achieve much higher reconstruction
accuracy than any other methods in the literature.  To apply this
method to observation, one needs to reconstruct the cosmic density
field of the local Universe. As illustrated in \citet{Wang_etal09},
this density field can be fairly well reconstructed using the
distribution of (relatively massive) galaxy groups (i.e., dark matter
halos). Using the galaxy groups \citep{Yang_etal07, Yang2012}
extracted from the SDSS Data Release 7
\citep[][DR7]{Abazajian_etal09}, \citet{Wang_etal12} have obtained the
mass, tensor and velocity fields of the local universe in the SDSS DR7
region.

With all these preparations, in a recent study, \citet[hereafter
ELUCID III]{Wang2016}, make use of the density field re-constructed
from the north galactic pole of SDSS DR7 region with an improved
domain mass assign method \citep{Wang_etal13} to predict the evolution
of structure of the local universe enclosed in a 500$\mpch$ length
cubic box.  As shown in ELUCID I, the reconstruction can recover more
than half of the phase information down to a scale $k\sim 3.4\hmpc$ at
$z=0$. Tests using original and reconstructed simulations show that
more than half of the halos with mass $\ga 10^{13.5}\msunh$ can be
reliably reproduced in which more than 50\% particles are in common
with the counterpart halos in the original simulation
\citep[see][hereafter ELUCID II]{Tweed2017}.  These features indicate
that the halos formed in the SDSS DR7 region in our ELUCID simulation
have roughly consistent large scale environment as the true universe
and the evolution of massive halos can be  roughly well modelled.

In this paper, we propose a novel neighborhood abundance matching
method to link galaxies observed in the SDSS DR7 region with
halos/subhalos in our ELUCID simulation {\it in the local same small
  volumes}. Once the galaxy-subhalo connections are generated, we can
use them to constrain semi-analytical galaxy formation models (SAMs)
in an halo based and local environment based apple-to-apple
comparisons. Note that, technically, we can make the neighborhood
abundance matching between galaxy groups and dark matter halos as
well. However, because of the following a few reasons, we decide to
use galaxies rather than groups. The main reason is that since some
massive groups in the observation may split (or connected) with
respect to halos in the simulation, then a group-halo matching may
over predict the galaxy population in one halo and under predict it in
the other. The second reason is that since only massive groups can be
well reproduced in the ELUCID simulation, we are not able to use the
group-halo connections in individual low mass halos.  In addition, it
would be interesting to see the impact of interlopers in the galaxy
groups which is not available in the group-halo matching
\citep[see][for the related discussions]{Campbell2015}.

The structure of this paper is organized as follows.  Section
\ref{sec_data} gives a detailed description of the data we used in
this study, including the halos/subhalos extracted from the ELUCID
simulations, the SDSS DR7 galaxy catalog, as well as the neighborhood
abundance matching between galaxies and dark matter subhalos.  In
Section~\ref{sec_hod} we probe the central-subhalo relations,
satellite fraction, CLFs and CSMFs of galaxies. In Section
~\ref{sec_bias} we measure the cross correlation functions between
galaxies and the dark matter particles. In Section \ref{sec_use} we
provide some suggestions for the usage of the galaxy-subhalo
connections established in this work.  Finally, we summarize our
results in Section~\ref{sec_conclusion}.  Throughout the paper we
adopt a $\Lambda$CDM cosmology with parameters that are consistent
with the fifth-year data release of the WMAP mission (hereafter WMAP5
cosmology): $\Omega_{\rm m} = 0.258$, $\Omega_{\Lambda} = 0.742$,
$\Omega_{\rm b} = 0.044$, $h=H_0/(100 \kmsmpc) = 0.72$ and
$\sigma_8 = 0.80$ \citep{Dunkley2009}.

\begin{figure*}
\plotone{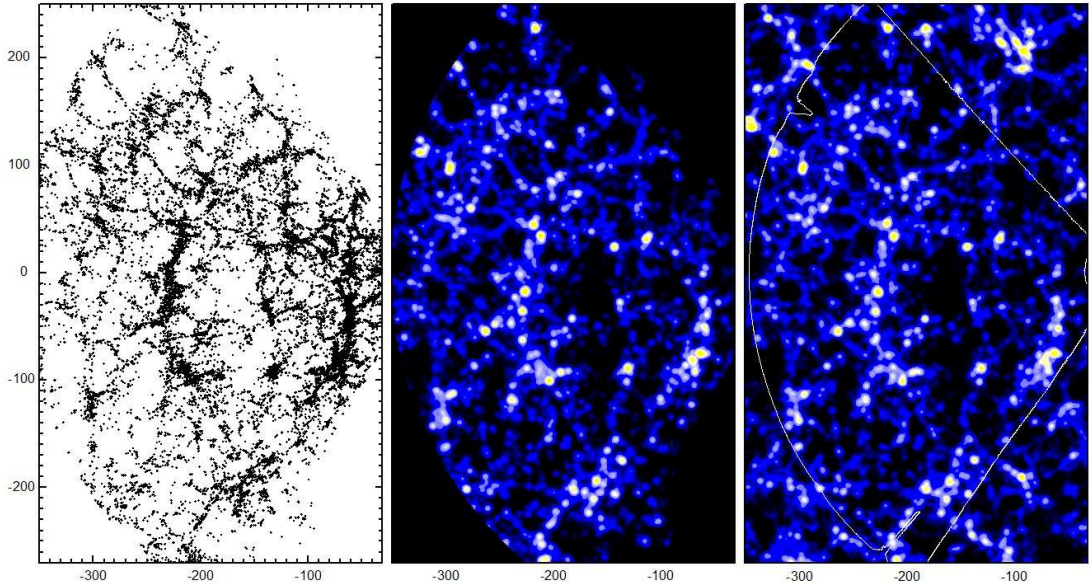}
\caption{The sketch of the SDSS DR7 galaxy and density fields in the
  ELUCID simulation. Shown in the left hand panel is a slice of galaxy
  distributions in the north galactic pole of SDSS DR7. In the middle
  panel is a slice of density field constructed from these galaxy
  distributions. In the right panel is the mass density field revealed
  in our ELUCID simulation. The enclosed region with solid lines is
  the density field supposed to be reproduced in the ELUCID
  simulation. }
\label{fig:ELUCID}
\end{figure*}

\section{data}
\label{sec_data}

\subsection{The halos/subhalos in the ELUCID simulation}
\label{sec:simulation}

In this study we use dark matter halos/subhalos extracted from the
ELUCID simulation. This simulation which evolves the distribution of
$3072^{3}$ dark matter particles in a periodic box of $500 \mpch$ on a
side was carried out in the Center for High Performance Computing,
Shanghai Jiao Tong University. The simulation was run with {\tt
  L-GADGET}, a memory-optimized version of {\tt GADGET2}
\citep{Springel2005b}. The cosmological parameters adopted by this
simulation are consistent with WMAP5 results with each particle has a
mass of $3.0875\times10^{8}\msunh$. In our ELUCID simulation, we make
use of the mass density field extracted from the galaxy/group
distribution in the north galactic pole region of the SDSS DR7 to
constrain the initial conditions using a Hamiltonian Markov Chain
Monte Carlo method (HMCMC) with particle mesh dynamics (see ELUCID I
\& III for details).  As an illustration, we show in
Fig. \ref{fig:ELUCID} the distributions of the related galaxies,
reconstructed and simulated mass density fields.  Shown in the left
hand panel is a slice of galaxy distributions in the north galactic
pole of SDSS DR7. In the middle panel is a slice of density field
constructed from these galaxy distributions. In the right panel is the
mass density field revealed (evolved to redshift $z=0$) in our ELUCID
simulation. The enclosed region in the left panel with solid lines is
the density field supposed to be reproduced in our ELUCID simulation.
The basic properties of our ELUCID simulation, including the algorithm
to perform the simulation, as well as the output power spectrum, halo
mass functions, etc. can be found in ELUCID III.

From the ELUCID simulation, dark matter halos were first identified by
a friends-of-friends (FOF) algorithm with linking length of $0.2$
times the mean particle separation and containing at least $20$
particles.  The dark matter halo mass function of this simulation at
redshift $z=0$ is checked and agrees very well with the model
predictions of \citet{Sheth2001} and \citet{Tinker2008}.  Based on
halos at different outputs, we first use the SUNFIND algorithm
\citep{Springel2001} to identify the bound substructures in the FOF
halos. The most massive substructure in a FOF halo is considered as
the main halo of this FOF and all the other subhalos in this FOF are
called subhalos.  For a given subhalo or main halo, each particle is
assigned a weight which decreases with the binding energy. We then
find all main halos and subhalos in the subsequent snapshot that
contain some of its particles. We count these particles with weight
for these potential descendants. The candidate with highest weighted
count is selected as the descendant. Please see \citet{Springel2005a}
for the details.

In order to properly link galaxies with dark matter halos and
subhalos, especially those satellite galaxies, one needs to properly
treat the subhalos in the simulations \citep[see][for the related
discussions]{Jiang2016}.  A widely adopted subhalo population in SHAMs
are the mass or circular velocity of survived subhalos extracted by
sub-finders in the simulation, but their masses/velocities are updated
to the maximum values along their accretion histories
\citep[e.g.][]{Conroy2006, Hearin2013}.  This method implies that: (1)
each subhalo can form only one galaxy, (2) the central-host halo
relation does not evolve significantly with redshift, and (3)
satellite galaxies disrupt whenever subhalos can no longer be
identified in the simulation, either because of limiting mass
resolution or because the subhalo is tidally disrupted. An alternative
method in the SHAM is to separate the central and satellite galaxies,
i.e., one can make abundance matching separately for central galaxies
v.s. main halos and satellite galaxies v.s. subhalos \citep[see][for a
similar attempt]{Rodriguez2015}.  In this regard, the evolution of
satellite galaxy will be automatically taken into account.  In this
study, we will perform both of these two kinds of abundance matching
methods and compare the differences between them. We refer the former
as `Match1' and the later as `Match2'.

In order to match the galaxies in the SDSS observations, we first
rotate the simulation box so that the re-simulated density field is in
superposition with the SDSS observation region. We then discard the
dark matter subhalos that are outside the survey sky coverage region
used for our density field re-construction. The $ra$, $dec$ and
$z_{\rm com}$ of the subhalos are calculated from their real space
positions. Then their final redshifts are obtained by adding the
peculiar velocities, with $z_{\rm obs} = z_{\rm com}+v_{\rm pec}/c$.
We trim subhalos within the redshift range $0<z<0.12$ for our
subsequent matching with galaxies in observations.

\subsection{SDSS DR7}
\label{sec:DR8}

The galaxy catalog we used for finding galaxy groups, making density
re-construction and performing the ELUCID simulation is the New York
University Value-Added Galaxy Catalog (NYU-VAGC; Blanton \etal 2005).
The catalog is compiled based on SDSS DR7 (Abazajian \etal 2009) but
with an independent set of significantly improved reductions.  From
the NYU-VAGC, we select all galaxies in the Main Galaxy Sample with an
extinction corrected apparent magnitude brighter than $r=17.72$, with
redshifts in the range $0.01 \leq z \leq 0.20$ and with a redshift
completeness ${\cal C}_z > 0.7$.  This gives a sample of $639,359$
galaxies with a sky coverage of 7748 square degrees.  In this study,
we make use of all the galaxies in this sample for our investigation.
Within these $639,359$ galaxies, $35,678$ do not have spectroscopic
redshifts, and are assigned with redshifts from their nearest
neighborhoods.

According to \citet[][hereafter Y07]{Yang_etal07}, the absolute
magnitudes of galaxies in bandpass $Q$ are computed using
\begin{equation}
\label{magn}
\Qmag = m_Q + \Delta m_Q - {\rm DM}(z) - K_Q - E_Q\,.
\end{equation}
Here ${\rm DM}(z) = 5 \log\left[D_L/(\mpch)\right] + 25$ is the
bolometric distance modulus calculated from the luminosity distance
$D_L$ using a WMAP5 cosmology.  $\Delta m_Q$ is the latest zero-point
correction for the apparent magnitudes, which converts the SDSS
magnitudes to the AB system, and for which we adopt
$\Delta m_Q = (-0.036,+0.012,+0.010,+0.028,+0.040)$ for
$Q=(u,g,r,i,z)$.  All absolute magnitudes are $K+E$ corrected to
$z=0.1$.  For the $K$ corrections we use the latest version of
`Kcorrect' (v4) described in Blanton \etal (2003; see also Blanton \&
Roweis 2007), which we apply to {\it all} galaxies that have
meaningful magnitudes and meaningful redshifts, including those that
have redshifts from alternative sources and those that have been
assigned the redshift of their nearest neighbor.  Finally, the
evolution corrections to $z=0.1$ are computed using
$E_Q = A_Q(z-0.1)$, with $A_Q = (-4.22,-2.04,-1.62,-1.61,-0.76)$ for
$Q=(u,g,r,i,z)$ (see Blanton \etal 2003).  Note that these evolution
corrections imply that galaxies were brighter in the past (at higher
redshifts).

\begin{figure*}
%\plotone{f2.eps}
\center
\vspace{0.5cm}
\includegraphics[height=15.0cm,width=13.0cm,angle=0] {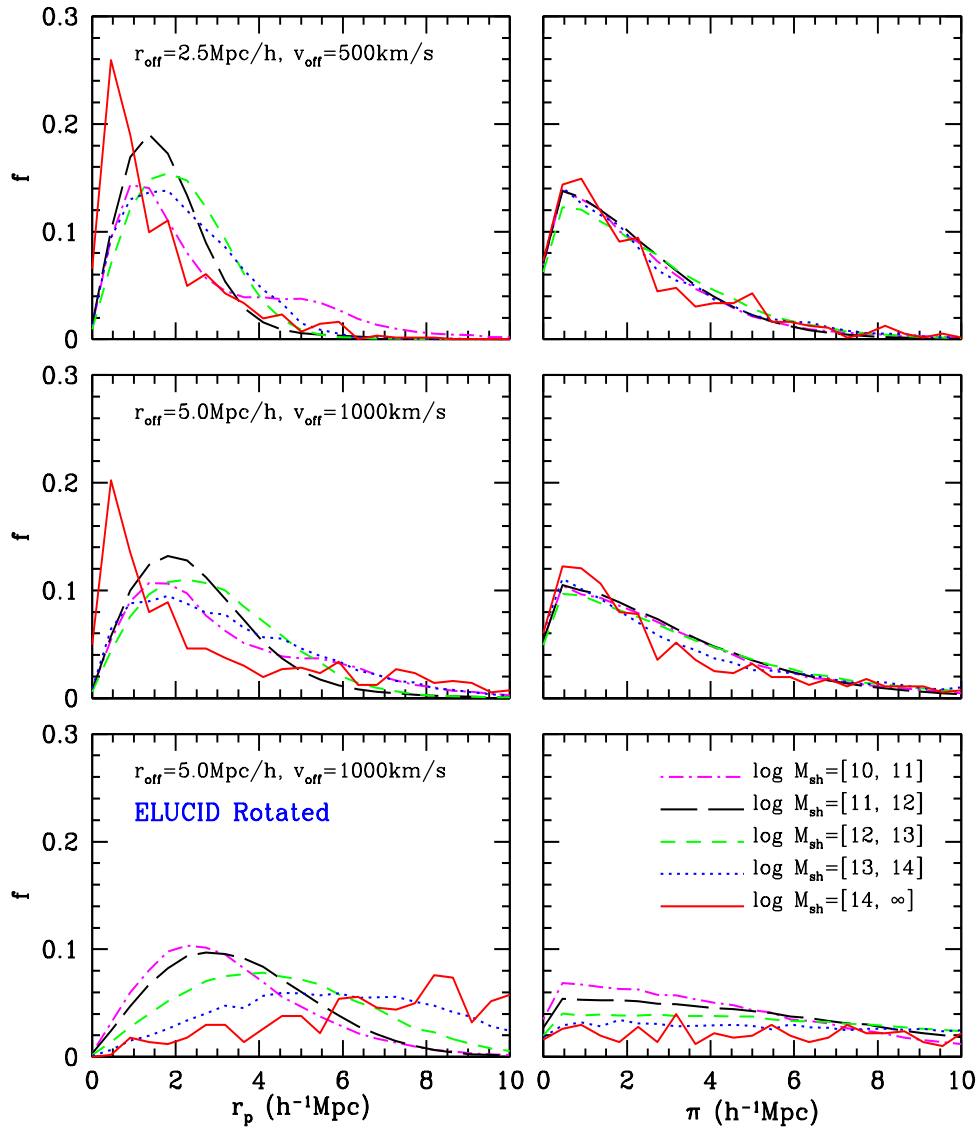}
\caption{The separation distributions of galaxy-subhalo pairs in
  different subhalo mass bins using different style lines as
  indicated. Here we only show results obtained from Match1
  method. Shown in the left and right panels are the $r_p$ and $\pi$
  distributions, respectively.  Results shown in the upper panels are
  obtained using criteria (1): $r_{\rm off}=2.5\mpch$ and
  $v_{\rm off}=500\kms$. Results shown in the middle panels are
  obtained using criteria (2): $r_{\rm off}=5.0\mpch$ and
  $v_{\rm off}=1000\kms$. While the results shown in the lower panels
  are obtained from a rotated (i.e. unmatched) version of the ELUCID
  simulation for matching criteria (2).  }
\label{fig:separation}
\end{figure*}
\begin{figure*}
%\plotone{f3.eps}
\center
\vspace{0.5cm}
\includegraphics[height=15.0cm,width=14.5cm,angle=0] {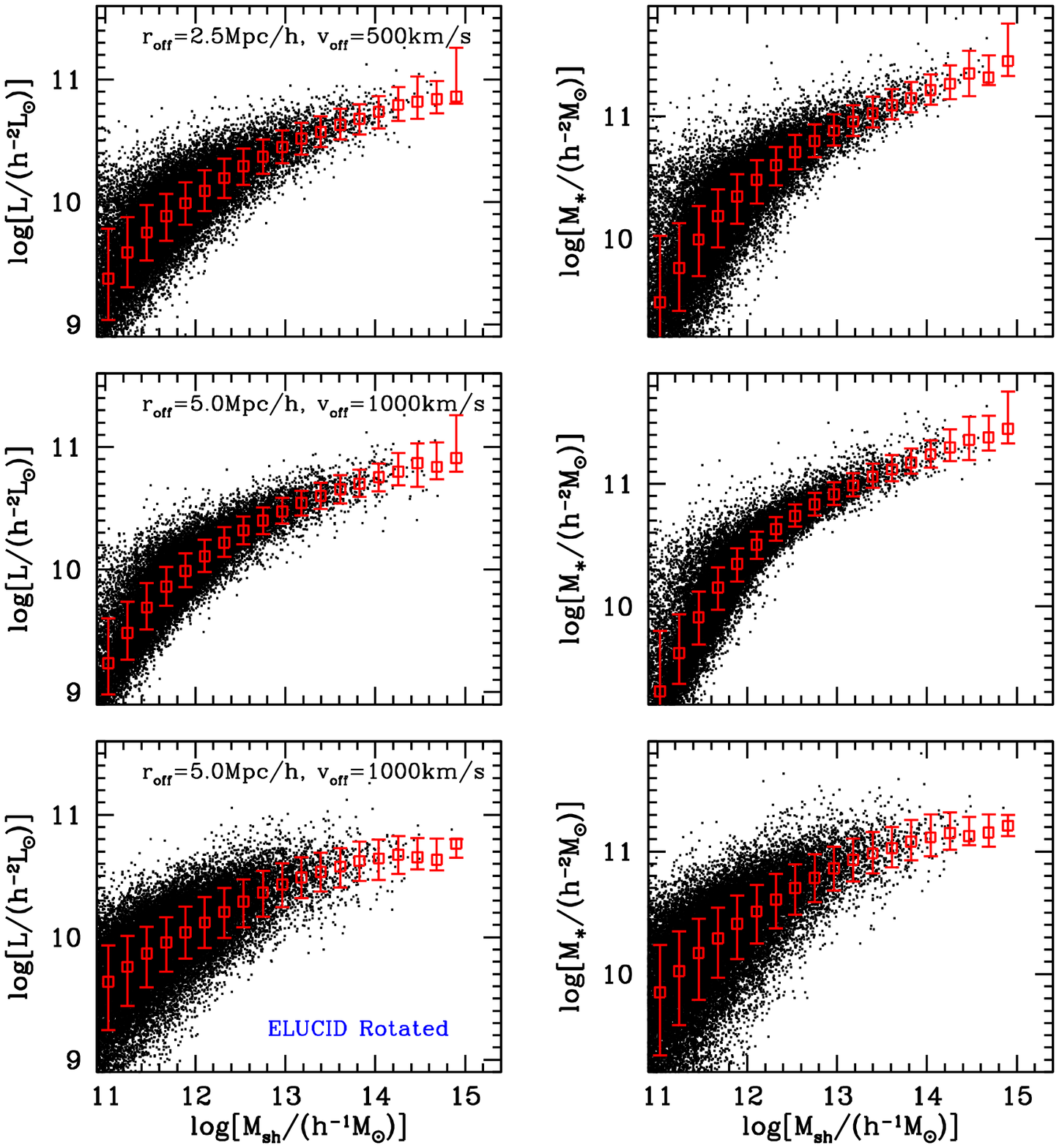}
\caption{The luminosity (left panels) and stellar mass (right panels)
  v.s. the subhalo mass $\log M_{sh}$ in the ELUCID simulation of the
  matched galaxy-subhalo pairs. Shown in the upper and middle panels
  are results obtained from Match1 method for matching criteria (1)
  and (2), respectively. While the results shown in the lower panels
  are obtained from a rotated (i.e. unmatched) version of the ELUCID
  simulation for matching criteria (2). }
\label{fig:Lc-M}
\end{figure*}

In addition to the absolute magnitudes, we also compute for each
galaxy its stellar mass, $M_*$. Using the relation between stellar
mass-to-light ratio and color of Bell \etal (2003), we obtain
\begin{eqnarray}
\label{eq:stellar}
\log\left[{M_* \over h^{-2}\Msun}\right] & = & -0.306 + 1.097
\left[^{0.0}(g-r)\right] - 0.1 \nonumber\\
& & - 0.4(\rrmag-4.64)\,.
\end{eqnarray}
Here $^{0.0}(g-r)$ and $\rrmag$ are the $(g-r)$ color and $r$-band
magnitude $K+E$ corrected to $z=0.0$, $4.64$ is the $r$-band magnitude
of the Sun in the AB system (Blanton \& Roweis 2007), and the $-0.10$
term effectively implies that we adopt a Kroupa (2001) IMF (Borch
\etal 2006).  For a small fraction (about $2\%$) of all galaxies, the
$g-r$ color that results from the photometric SDSS pipeline is
unreliable. These galaxies typically have $g-r$ colors that are
clearly unrealistic (they are catastrophic outliers in the
color-magnitude distribution).  If this is not accounted for,
equation~(\ref{eq:stellar}) assigns these galaxies stellar masses that
are unrealistically high or low. To take account of these outliers we
follow Y07 using the color bi-Gaussian distributions of galaxies
obtained in Yang et al. (2008; see also Li et al. 2006).  For any
galaxy that falls outside the 3-$\sigma$ ranges from the mean
color-magnitude relations of both the red sequence and the blue cloud
(about 2\% of all galaxies), we compute its stellar mass using the
mean color of the red sequence (when the galaxy is too red) or the
blue cloud (when the galaxy is too blue).

In order to probe the color dependence of galaxies, following Yang et
al. (2008), we separate our galaxies with $^{0.1}(g-r)<0.9$ into red
and blue subsamples using the criteria,
\begin{equation}\label{quadfit}
^{0.1}(g-r) = 1.022-0.0651x-0.00311x^2\,,
\end{equation}
where $x=\rmag + 23.0$. While galaxies with $^{0.1}(g-r)\ge 0.9$ are
directly attributed to red subsample.

From the above galaxy catalog, we only select galaxies within groups
that were used to map the density field and thus the initial density
field in our ELUCID simulation. That is we only select galaxies which
are located within the range: $99<ra<283$, $ -7<dec<75$ (i.e. in the
north galatic pole) and redshift $0.01<z<0.12$.  After this selection,
a total of 396069 galaxies are remained for our subsequent probes.

\begin{figure*}
%\plotone{f3.eps}
\center
\vspace{0.5cm}
\includegraphics[height=13.0cm,width=14.5cm,angle=0] {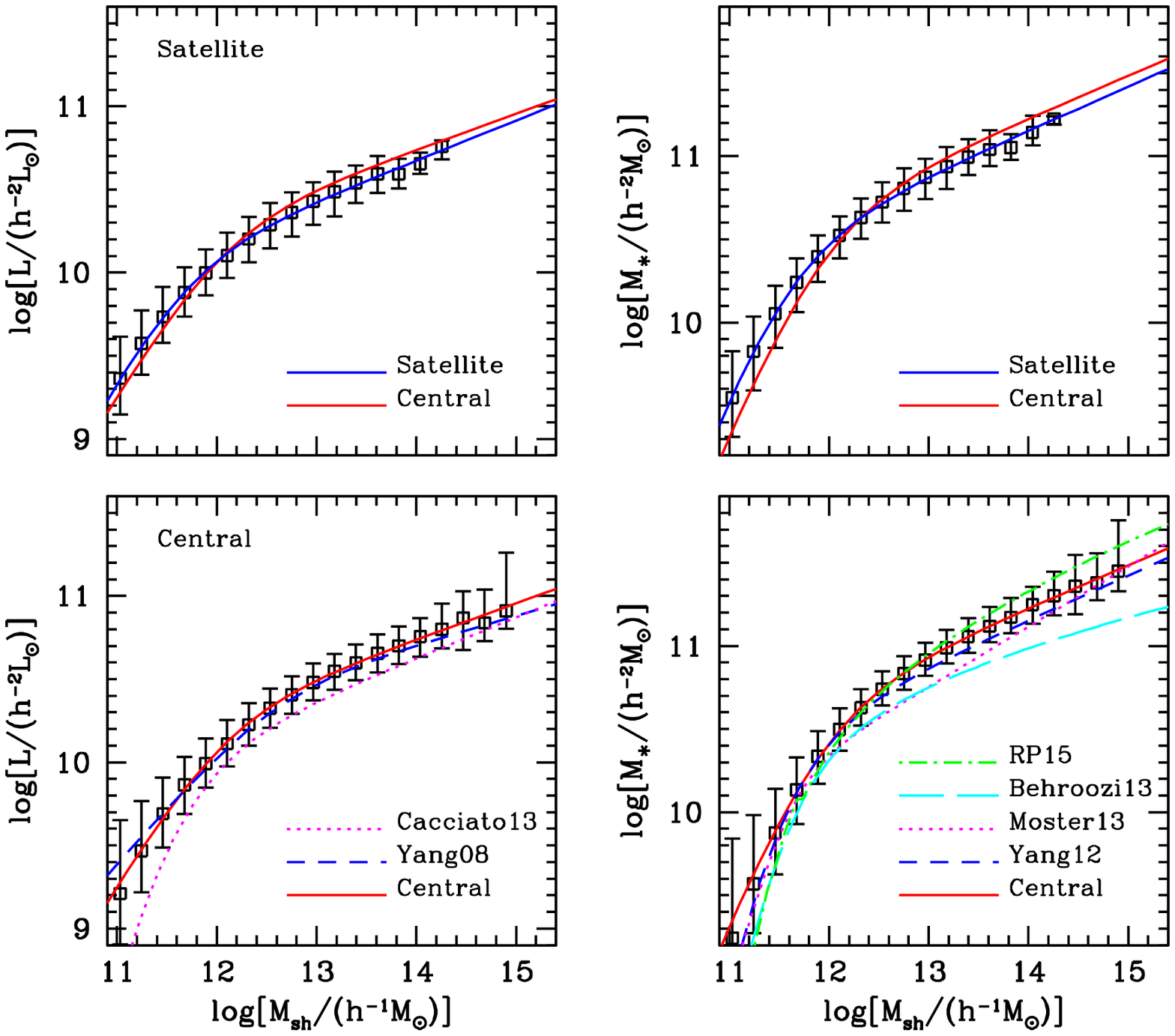}
\caption{The luminosity (left panels) and stellar mass (right panels)
  v.s. the subhalo mass $\log M_h$ in the ELUCID simulation of the
  matched galaxy-subhalo pairs. Shown in the upper and lower panels
  are results obtained from Match2 method for central and satellite
  galaxies, respectively. The red and blue solid lines in these panels
  are the best fitting results for central and satellite galaxies
  separately.  For comparison, the blue dashed and magenta dotted
  dashed lines in the left panel are results obtained by
  \citet{Yang2008} and \citet{Cacciato2013}, respectively.  The blue
  dashed, magenta dotted, cyan long dashed and green dot-dashed lines
  are results obtained by \citet{Yang2012}, \citet{Moster2013},
  \citet{Behroozi2013} and \citet{Rodriguez2015}, respectively.  }
\label{fig:Lc-M2}
\end{figure*}

\subsection{The neighborhood abundance matching method}
\label{sec:algorithm}

In order to link galaxies with dark matter (sub)halos, one can either
establish the HOD or CLF models \citep[e.g.][]{ Jing1998, Peacock2000,
  Yang2003, Bosch2003, Zheng2005, Tinker2005, Cooray2006, Bosch2007,
  Brown2008, More2009, Cacciato2009, Neistein2011, Avila2011,
  Leauthaud2011, Yang2012, Rodriguez2015, Li2016, Zu2016, Bull2017,
  Cohn2017, Contreras2017, Rodriguez2017}, or via subhalo abundance
matching processes \citep[e.g.][]{Vale2004, Vale2006, Conroy2006,
  Shankar2006, Conroy2009, Moster2010, Guo2010, Behroozi2010,
  Hearin2013}.  These probes have revealed many observational features
of galaxies, and were widely used to constrain the galaxies formation
models. In these studies, however, only the global properties of
galaxies such as the stellar mass functions/luminosity functions, and
clusterings are used to establish the galaxy-(sub)halo connections.
In this probe, as the structures in our ELUCID simulation are supposed
to trace the evolution of real structures in our SDSS DR7 region in
consideration, we set out to match galaxies with the dark matter
subhalos in their neighborhood, i.e., using a neighborhood abundance
matching method.

We first sort the stellar masses of the galaxies.  Starting from the
most massive galaxy, we search in redshift space for each subhalo
sample the most likely subhalo in a small volume in its neighborhood,
which is then marked as its counterpart. The likelihood of the subhalo
to be linked with the candidate galaxy is modelled as follows,
\begin{equation}
P(r_p,\pi, M_{sh}) = M_{sh} \exp(- \frac{
  r_p^2} {2 r_{\rm off}^2}) \exp(- \frac{\pi^2} {2 v_{\rm off}^2})\,.
\end{equation}
Here $r_p$ and $\pi$ are the separations between the galaxy and
subhalo in the perpendicular and along the line of sight directions,
respectively.  $M_h$ is the mass of subhalo in consideration. While
$r_{\rm off}$ and $v_{\rm off}$ are the two free parameters we choose
to make our neighborhood abundance matching.  In the extreme case
where $r_{\rm off}=\infty$ and $v_{\rm off}=\infty$, the neighborhood
abundance matching method degrades to the traditional abundance
matching method.  We use two sets of parameters to perform our
neighborhood abundance matching: (1) $r_{\rm off}=2.5\mpch$ and
$ v_{\rm off}= 500\kms$ and (2) $r_{\rm off}=5.0\mpch$ and
$ v_{\rm off}= 1000\kms$, and compare their performances.  Note that
in our ELUCID simulation, the reconstructed density field is smoothed
using a Gaussian kernel of radius $2.0\mpch$. Here, these two sets of
choices are made according to a compromise of the scatters and the
separations between the galaxies and subhalos, which will be
illustrated as follows.  Using these criteria, we sequentially search
for all the galaxies their counterparts in redshift space within a
maximum distance $\le 30\mpch$.  For the total of 396069 galaxies,
according to criteria (2), there are 296488 central galaxies that are
linked with the main halos and 99581 satellite galaxies that are
linked with the subhalos for Match1 method.  Criteria (1) gives very
similar numbers, with typical differences at a few hundreds. Comparing
to the ones specified in the group catalog, 277139 centrals and 118930
satellites, the Match1 method roughly underestimated about $\sim 20\%$
satellite galaxy population. On the other hand, by definition, the
Match2 method will give the same central/satellite separation as those
in groups.

We show in Fig. \ref{fig:separation} the separation distributions of
galaxy-subhalo pairs in different subhalo mass bins. Shown in the
upper-left panel are the $r_p$ distributions for matching criteria
(1): $r_{\rm off}=2.5\mpch$ and $v_{\rm off}=500\kms$. Note that as
the results for Match1 and Match2 methods are very similar, here we
only show those obtained from Match1 method. As we can see, the offset
between galaxies and subhalos in the most massive mass bin with
$\log M_h\ge 14.0$, which peaks at $\sim 0.5 \mpch$, is the
smallest. About 50\% of the matched pairs have projected separations
less than $2\mpch$. The lower mass subhalos have slightly larger
separations and the distributions peak at about $1.5-2.0 \mpch$.
About 50\% of the matched pairs have projected separations less than
$2.5\mpch$. Shown in the upper-right panel are the $\pi$ distributions
of galaxy-subhalo pairs for matching criteria (1). The offsets for
subhalos in different mass bins are quite similar. All the
distributions peak at $\sim 50\kms$ and about 50\% of the matched
pairs have line of sight separation less than $200\kms$.

The results shown in the middle panels of Fig. \ref{fig:separation} are
similar to those shown in the upper panels, but for matching criteria
(2): $r_{\rm off}=5.0\mpch$ and $v_{\rm off}=1000\kms$. The overall
distribution properties are quite similar to those of matching
criteria (1), but with slightly larger offsets. That is, about 50\% of
the matched pairs have projected separations less than $3.0\mpch$ and
line of sight separation less than $250\kms$.

\begin{figure*}
%\plotone{f4.eps}
\center
\vspace{0.5cm}
\includegraphics[height=13.0cm,width=13.0cm,angle=0] {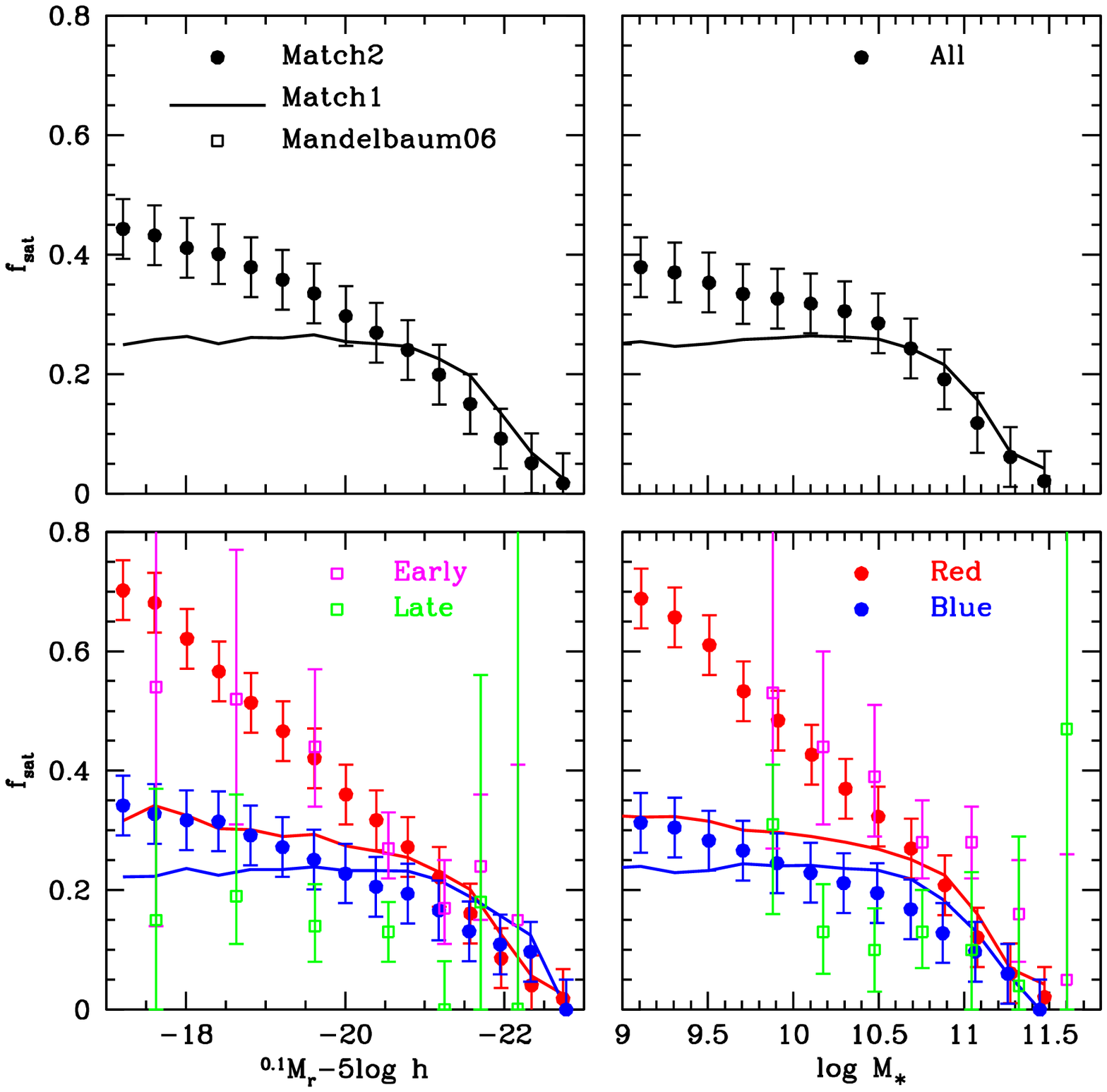}
\caption{Fraction of satellite galaxies as function of luminosity
  (left panels) and stellar mass (right panels). Results are shown
  separately for all (upper panels), red and blue galaxies (lower
  panels), respectively. The solid dots with error bars represent the
  results obtained for Match2 method. Lines are results for Match1
  method. The open squares with error bars are results obtained by
  \citet{Mandelbaum2006} for comparison.  }
\label{fig:f_sat}
\end{figure*}
\begin{figure*}
\plotone{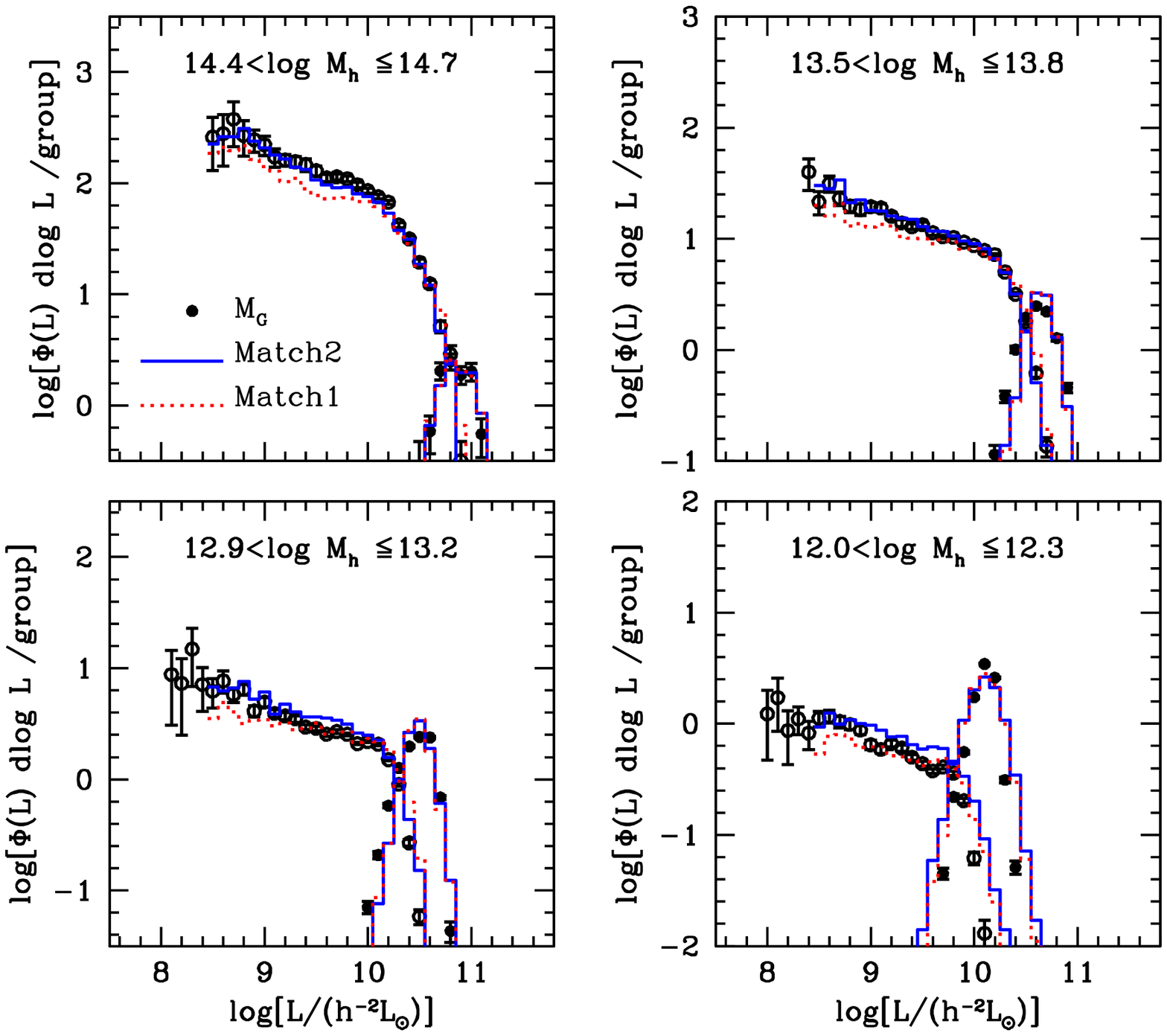}
\caption{The conditional luminosity functions (CLFs) of galaxies in
  groups and host halos of different masses.  Symbols correspond to
  the CLFs obtained from SDSS galaxy groups, with solid and open
  circles indicating the contributions from central and satellite
  galaxies, respectively. The errorbars reflect the 1-$\sigma$ scatter
  obtained from 200 bootstrap samples.  The solid and dotted lines are
  results obtained for the Match2 and Match1 methods, respectively.}
\label{fig:CLF}
\end{figure*}

As we have matched galaxies with subhalos, it is quite straightforward
to check their luminosity (stellar mass) - subhalo mass relations.
Show in the upper panels of Fig.~\ref{fig:Lc-M} are the luminosity -
subhalo mass (left panel) and stellar mass - subhalo mass (right
panel) relations for our matching criteria (1). Shown in the middle
panels are results for our matching criteria (2).  Here again, we only
show results obtained from Match1 method. While results obtained from
Match2 are very similar. In each panel, the open squares with
error-bars indicate the median and 68\% confidence levels of these
relations of all the galaxies. Comparing the results for the two
matching criteria, the latter shows somewhat tighter luminosity
(stellar mass) - subhalo mass relations, especially in the low mass
subhalo. We thus think the latter matching criteria works better. In
what follows, we only present results obtained using the matching
criteria (2), i.e., with $r_{\rm off}=5.0\mpch$ and
$v_{\rm off}=1000\kms$.

Before we proceed to provide more detailed tests on the performance of
our neighborhood abundance matching method on the ELUCID simulation,
it would be interesting to check the above separation distribution and
luminosity (stellar mass) - subhalo mass relations if the neighborhood
abundance matching method is applied to a simulation that does not
have good correspondence with the observation. For this purpose, we
rotate the ELUCID simulation box by 90 degree and shift it by
$250\mpch$, and then perform the same procedures using matching
criteria (2).  Note that after such a treatment, the simulation
density field is no longer matched with the SDSS density field. Shown
in the lower panels of Figs. \ref{fig:separation} and \ref{fig:Lc-M}
are the resulting separation distributions and luminosity (stellar
mass) - subhalo mass relations. We can see that the separation
distributions of galaxy-subhalo pairs in this situation are very
different from our fiducial case, especially for massive clusters.
The very large separation between galaxies and massive (sub)halos
indicates that the galaxy-subhalo pairs might come from different
origins.  For small (sub)halos with mass $\la 10^{12.0}\msunh$, the
difference is quite small, indicating that the low mass galaxy-subhalo
pairs, even in the ELUCID simulation might dominated by Poisson errors
\citep[see][]{Tweed2017}. In addition, the luminosity (stellar mass) -
subhalo mass relations in this case are much worse, i.e., with much
larger scatters, than those of our fiducial cases.  In general, as we
mentioned, if we set $r_{\rm off}=\infty$ and $v_{\rm off}=\infty$,
the neighborhood abundance matching method will degrade to the
traditional abundance matching method, which will provide monotonic
luminosity (stellar mass) - subhalo mass relations.

\section{The halo occupation distributions of galaxies}
\label{sec_hod}

\begin{figure*}
  \plotone{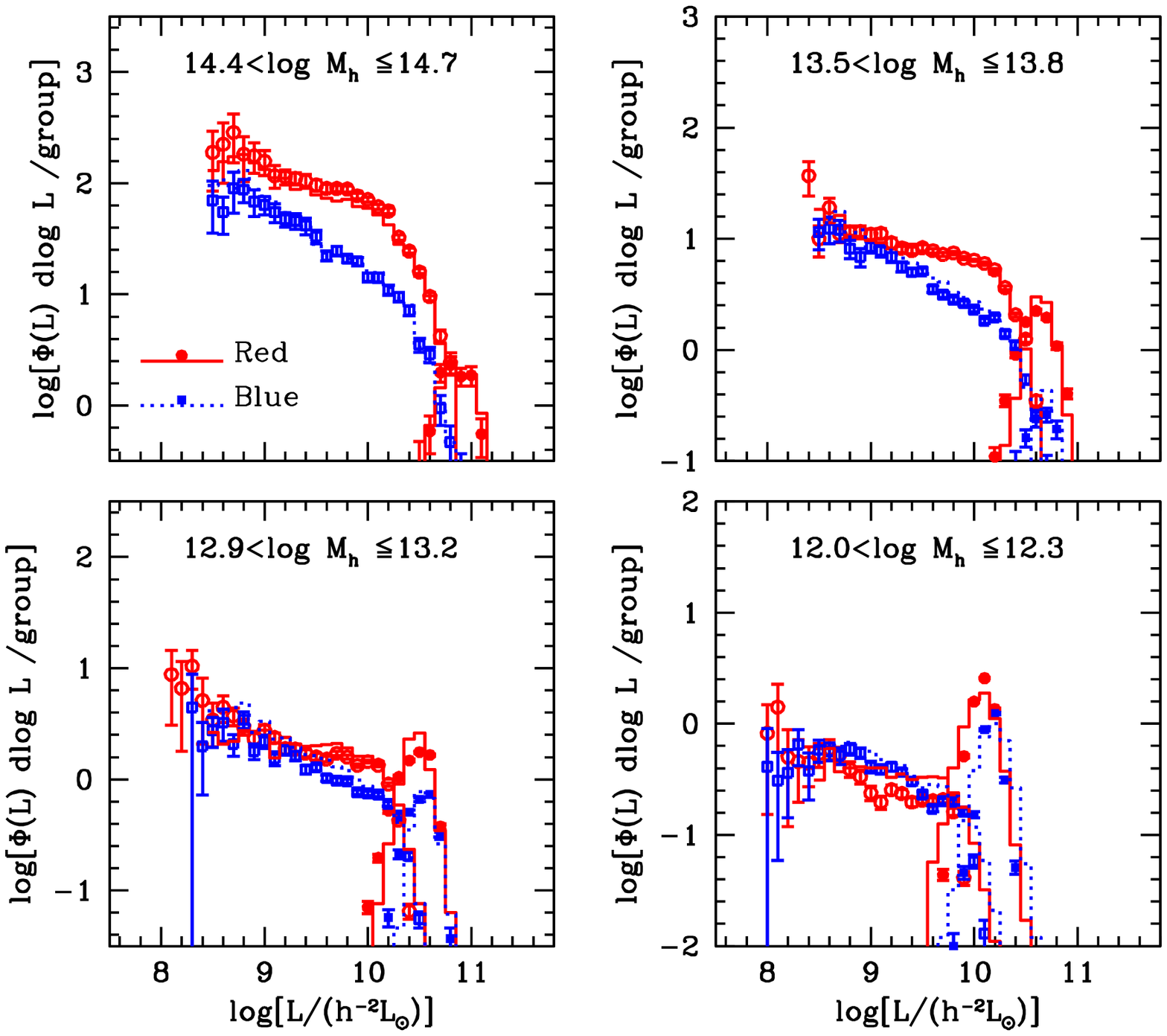}
  \caption{Similar to Fig.~\ref{fig:CLF}, but here we show the CLFs
    for red and blue galaxies.  In both cases the central and
    satellite components of the CLFs are indicated separately.  Here
    symbols and lines represent results obtained from SDSS galaxy
    groups and Match2 method, respectively. }
\label{fig:CLF_color}
\end{figure*}

After we matched galaxies with subhalos in the ELUCID simulation, we
proceed to measure a few galaxy statistics within host halos of
different masses. These statistics are compared to those obtained from
galaxy groups \citep[e.g.][]{Yang2008, Yang2009} to demonstrate the
feasibility of populating dark matter subhalos with observed galaxies
via the neighborhood abundance matching method outlined in 
section \ref{sec:algorithm}.

\subsection{The galaxy-subhalo luminosity/mass relations}

Here we start our probe using the data obtained from Match2 method.
Shown in the upper and lower panels of Fig. \ref{fig:Lc-M2} are the
luminosity (stellar mass) - subhalo mass relations for satellite and
central galaxies separately.  We follow \cite{Yang2008} to use the
following $L-M_{sh}$ functional form to describe the median luminosity
- subhalo mass relation,
\begin{equation}\label{eq:Lc_fit}
L = L_0 \frac { (M_{sh}/M_1)^{\alpha +\beta} }{(1+M_{sh}/M_1)^\beta } \,.
\end{equation}
This model contains four free parameters: a normalized luminosity,
$L_0$, a characteristic halo mass, $M_1$, and two slopes, $\alpha$ and
$\beta$. The blue and red solid lines shown in the left panels are the
best fits to the average $L-M_{sh}$ relations for satellite and
central galaxies, respectively.  The best fitting parameters are
listed in Table \ref{tab:tbl-1} in the first and second rows.
Although not very significant, we do see some differences between the
luminosity - subhalo mass relations of central and satellite galaxies,
indicating that satellite galaxies may have different stripping or
disruption effect from subhalos.

For comparison, we also show in the lower-left panel the best fitting
results obtained by \citet{Yang2008} using a dashed line, where the
set of best fitting parameters are listed in the third row of Table
\ref{tab:tbl-1}.  This set of result is obtained from SDSS galaxy
group catalog directly, in which (i) it is assumed that central
galaxies are the brightest group members, and (ii) halo masses are
inferred using abundance matching of host halos to the total stellar
mass of the groups. The dotted line shown in that panel are the
results obtained by \citet{Cacciato2013} by CLF model constraints
using the clustering and weak lensing measurements of
galaxies. Overall, our neighborhood abundance matching method gives
quite consistent $L-M_{sh}$ relation with these previous measurements,
except that of \citet{Cacciato2013}. The slight systematic deviation
from \citet{Cacciato2013} is mainly caused by the different definition
of halo mass and the cosmology they used. As shown in Fig. 7 of
\citet{Cacciato2013}, if the halo mass definition and cosmology are
properly converted, their results agree with those obtained by
\citet{Yang2008} very well.

For the $M_{*}-M_{sh}$ relations shown in the right panels, we use a
similar function to fit the data:
\begin{equation}\label{eq:Mc_fit}
M_{*} = M_0 \frac { (M_{sh}/M_1)^{\alpha +\beta} }{(1+M_{sh}/M_1)^\beta } \,.
\end{equation}
The blue and red solid lines shown in the right panels are the best
fits of this model to the data for the satellite and central galaxies
separately, where the best-fit parameters are listed in the fourth and
fifth rows of Table \ref{tab:tbl-1}.

For comparison, we also show in the lower-right panel, the model
constraints obtained by \citet{Yang2012} using a dashed line, where
the related set of parameters are listed in the sixth row of Table
\ref{tab:tbl-1}.  The model constraints obtained by
\citet{Moster2013}, \citet{Behroozi2013} and \citet{Rodriguez2015} are
shown as the dotted, long dashed and dot-dashed lines, respectively.
Here again, we see our neighborhood abundance matching method gives
quite consistent $M_{*}-M_{sh}$ relation with these previous probes,
except that of \citet{Behroozi2013}, which is somewhat lower
especially at massive end.  The difference is mainly caused by
adopting a different stellar mass estimation method \citep[see][for a
similar trend and the related discussions]{Behroozi2013}.
 
\begin{table}[h!]
\begin{center}
  \caption{\label{tab:tbl-1} The best fitting parameters.}
\begin{tabular}{lcccc}
  \hline\hline
  Sample & $\log L_0$ &  $\log M_1$ & $\alpha$& $\beta$\\
  \hline
  Satellite  & 10.093 &  11.570  & 0.240  &  0.936 \\
  Central  &  10.316  & 12.024   & 0.215 &  0.795 \\
  Yang08  & 10.45 &   12.54 &   0.175 &    0.514 \\
\hline
\\
\hline \hline 
Sample & $\log M_0$&  $\log M_1$ & $\alpha$& $\beta$\\
\hline
  Satellite & 10.477   & 11.449 &   0.265 &   1.448 \\
  Central  &  10.680  & 11.875  &  0.257  &  1.236    \\
  Yang12 & 10.36 &  11.06 &  0.27 & 4.34 \\
  \hline
\end{tabular}
\end{center}
\end{table}

\subsection{The satellite fraction}

The second quantity we probe is the satellite fraction of
galaxies. Since a satellite galaxy resides in a more massive halo than
a central galaxy of the same luminosity or stellar mass
\citep[e.g.][]{Yang2003}, thus the fraction of satellite galaxies as
function of luminosity, $f_{\rm sat}(L)$, or stellar mass,
$f_{\rm sat}(M_{*})$, plays an important role in modelling both the
small and large scale clustering of galaxies of a given
luminosity/stellar mass \citep[e.g.][]{Jing1998, Berlind2002,
  Yang2003, Bosch2007, Yang2012}.  The satellite fraction as function
of luminosity, $f_{\rm sat}(L)$, is also important for a proper
interpretation of the measurements of galaxy-galaxy lensing signals
\citep[e.g.][]{Guzik2002, Mandelbaum2006, Yang2006} and pairwise
velocity dispersion of galaxies \citep[e.g.][]{Jing2004, Yang2004},
and to understand the quenching of galaxies \citep[e.g.][]{Bluck2016,
  Wang2017}.

\begin{figure*} \plotone{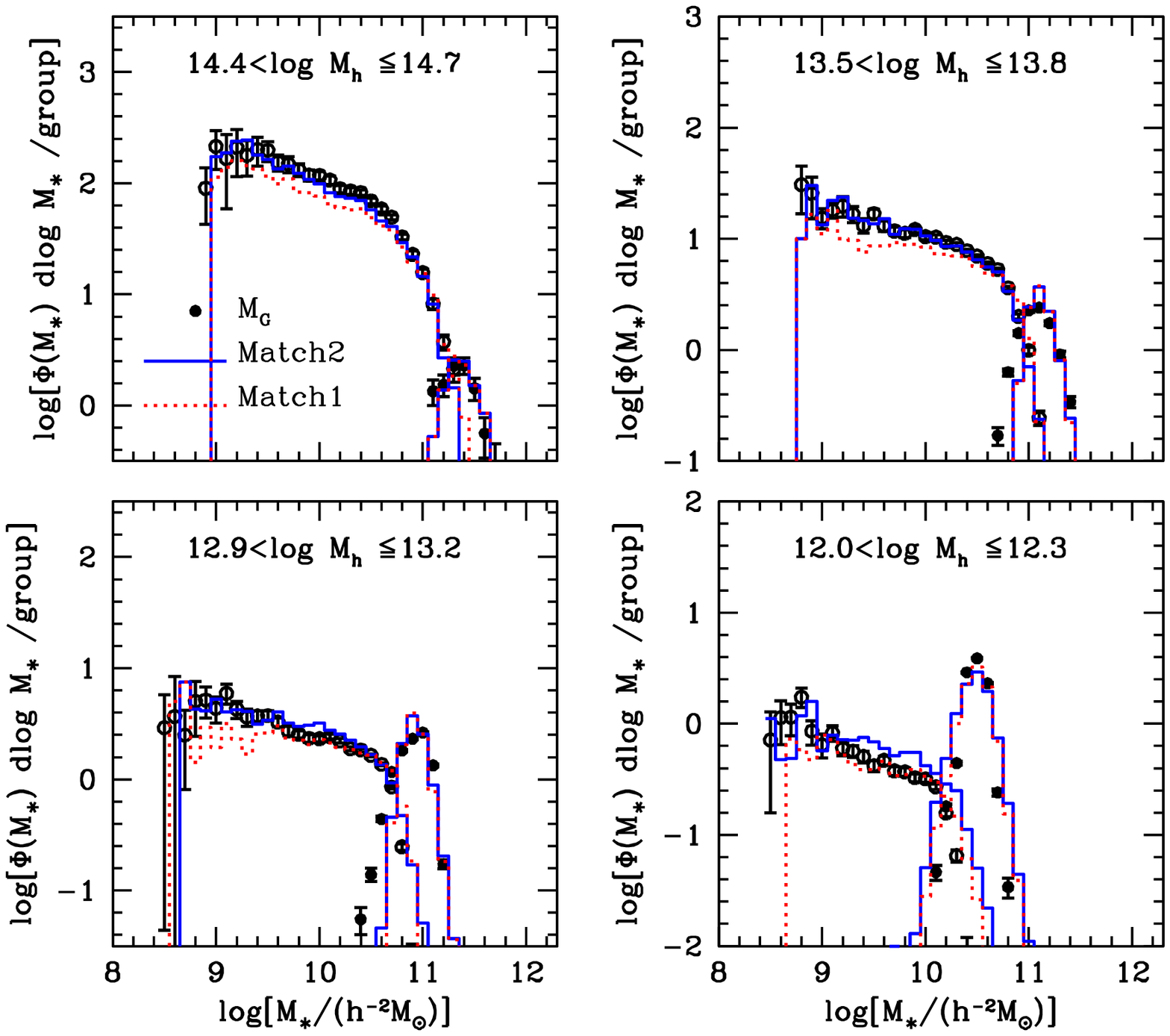}
  \caption{The conditional stellar mass functions (CSMFs) of galaxies
    in groups and host halos of different masses.  Symbols correspond
    to the CSMFs obtained from SDSS galaxy groups, with solid and open
    circles indicating the contributions from central and satellite
    galaxies, respectively.  The errorbars reflect the 1-$\sigma$
    scatter obtained from 200 bootstrap samples.  The solid and dotted
    lines are results obtained for the Match2 and Match1 methods,
    respectively.}
\label{fig:CSMF}
\end{figure*}
\begin{figure*} \plotone{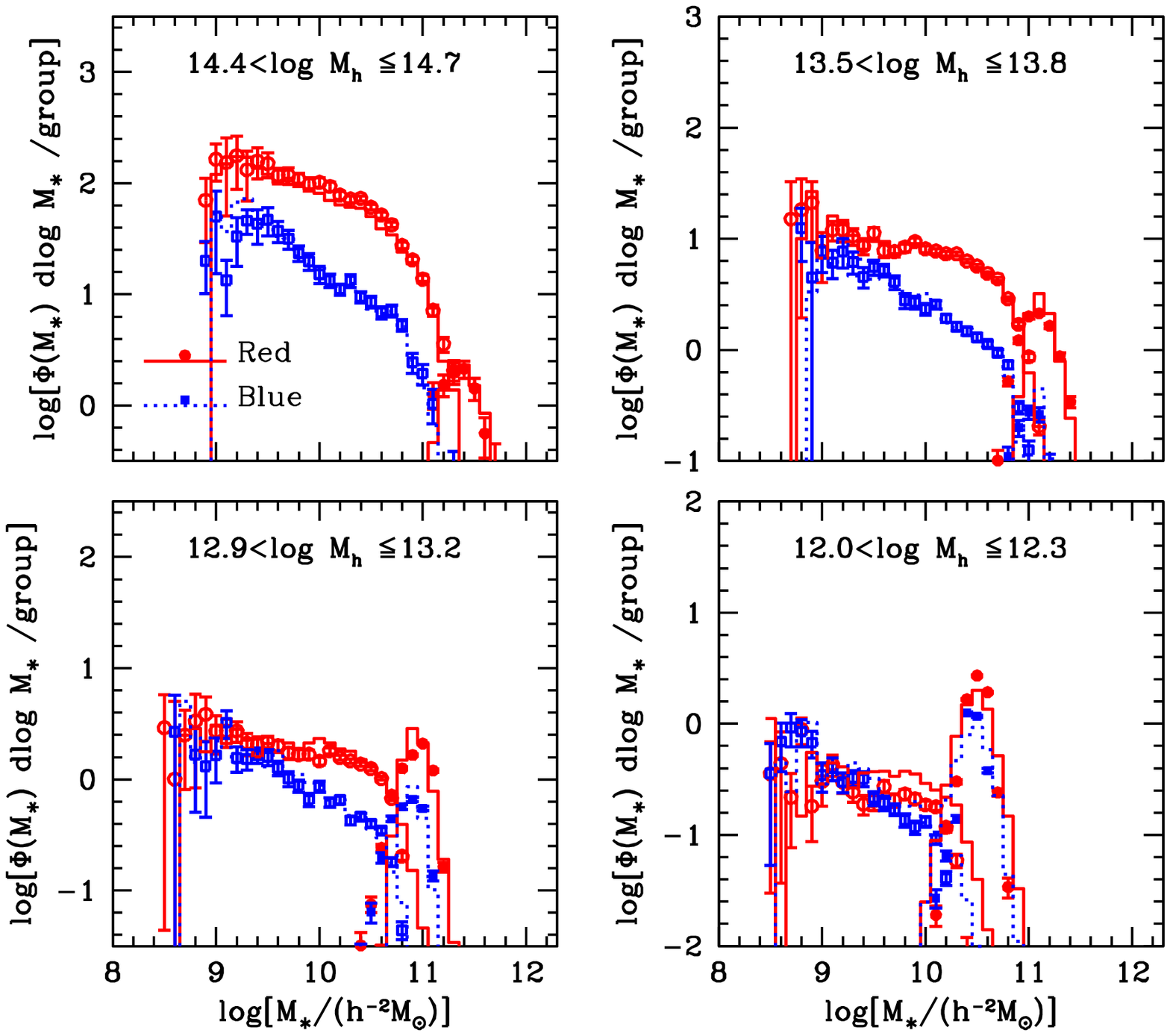}
  \caption{Similar to Fig.~\ref{fig:CSMF}, but here we show the CSMFs
    for red and blue galaxies. Here symbols and lines represent
    results obtained from SDSS galaxy groups and Match2 method,
    respectively.}
\label{fig:CSMF_color}
\end{figure*}

Here we estimate $f_{\rm sat}(L)$ and $f_{\rm sat}(M_{*})$ directly
from our matched galaxy-subhalo pairs.  In the left-hand panels of
Fig.~\ref{fig:f_sat} we show $f_{\rm sat}(L)$ as a function of galaxy
luminosity. The results are plotted separately for all (upper panels),
red and blue (lower panels) galaxies.  Since in our Match2 method the
central and satellite galaxies are matched with main halos and
subhalos separately, by definition, the satellite fractions in our
Match2 method follow the same as those of SDSS galaxy groups. We show
the resulting satellite fraction in Fig.~\ref{fig:f_sat} using solid
dots with error bars. Note that, in our probe, we have made use of the
modelC sample in \citet{Yang_etal07}, where about 5\% of galaxies that
lacking spectroscopic redshifts due to fiber collision effect are
assigned with redshifts from their nearest neighbors. As pointed out
in \citet{Yang2008}, fiber collisions are expected to significantly
impact the number of close pairs and hence the satellite fractions
$f_{\rm sat} (L)$.  The typical uncertainties induced by adding or
removing the fiber collision galaxies are about 5\%. Here we adopt
this uncertainty value as the error bars shown in
Fig.~\ref{fig:f_sat}.

First for all galaxies, by comparing the model predictions of the
Match1 method with those of Match2 method, we can see, Match1 method
predicts roughly consistent satellite fractions for relatively bright
galaxies.  However, if one goes to fainter galaxies with
$\rmag >-20.0$, the satellite fractions are significantly
underestimated. This discrepancy indicates that the widely used
subhalo abundance matching method in literature may not predict the
low mass satellite galaxies correctly. One can either add more
subhalos (e.g., the disrupted subhalos) in their abundance matching
with galaxies, or match central and satellite galaxies separately as
we did here in Match2 method.

Next for galaxies that are separated into red and blue populations,
compare to those measured from the galaxy groups or Match2 method, the
model predictions for Match1 method show much smaller segration
between red and blue galaxies.  Note that since in our neighborhood
abundance matching procedures, we did not make any special treatments
between red and blue galaxies, thus the lack of segration for Match1
method is somewhat expected.  In general, one may treat red and blue
galaxies differently to have a better model prediction of red/blue
satellite fractions \citep[e.g.][]{Rodriguez2015}, or more
straightforwardly by matching central and satellite galaxies
separately.  

The satellite fraction as a function of galaxy stellar mass are shown
in the right panels of Fig.~\ref{fig:f_sat}.  The overall behaviors
are quite similar to those shown in the left panels. 

Finally, as a comparison, we also show in the lower panels of
Fig. \ref{fig:f_sat} the satellite fraction obtained by
\citet{Mandelbaum2006} for early and late type galaxies
(open squares with $95\%$ confidence level error bars) from the
galaxy-galaxy weak lensing measurements. Although their samples are
defined differently from ours (early and late types according to
galaxy morphologies, v.s. red and blue galaxies according to color),
the two measurements agree very well.

\subsection{The conditional luminosity functions}

The conditional luminosity function (CLF) of galaxies in dark halos,
$\Phi(L \vert M)$, which describes the average number of galaxies as a
function of galaxy luminosity in a dark matter halo of a given mass,
plays an important role in our understanding of how galaxies form in
dark matter halos \citep[e.g.][and references therein]{Yang2003,
  Yang2012}.  Here we directly measure $\Phi(L \vert M)$ from our
matched galaxy-subhalo pairs, and compare them to those obtained from
the galaxy group catalogs. In order to make proper comparisons, we
updated the halo masses of galaxy groups according to the WMAP5
cosmology adopt in this study.

The CLF can be estimated by directly counting galaxies in halos and
groups.  However, since the galaxies used for our study are flux
limited to $r=17.72$, for a given galaxy luminosity $L$, there is a
limiting redshift, $z_L$, beyond which galaxies with such a luminosity
are not included in the sample. In order to estimate $\Phi(L \vert M)$
at a given $L$, we only use halos and groups that are complete to the
redshift limit $z_L$.  The CLF is obtained by simply counting the
average number of galaxies (in luminosity bins) in halos or groups of
a given $M$.  We show in Fig.~\ref{fig:CLF} the resulting CLFs
obtained from galaxy groups of different masses using symbols with
error bars, where the error bars are obtained using 200 bootstrap
re-samplings of the groups.  The contributions of central and
satellite galaxies are plotted separately using filled and open
symbols, respectively. The solid lines shown in Fig.~\ref{fig:CLF} are
the results obtained for our fiducial Match2 method. For comparison,
we also show using dotted lines the results obtained for the Match1
method. First, for the central galaxy component, we see that both
Match2 and Match1 methods give very similar predictions. According to
the comparisons with the data points alone, we see that both Match2
and Match1 methods only agree with data in the most massive bin. In
all other three halo mass bins, the CLFs of central galaxies show
significant deviations. On the other hand however, if we model the
CLFs for central galaxies with a lognormal distribution
\citep[e.g.][]{Yang2008},
\begin{equation}\label{eq:phi_c}
\Phi_{\rm cen}(L_c|M) = {1\over {\sqrt{2\pi}\sigma_c}} {\rm exp}
\left[- { {(\log L_c  -\log L )^2 } \over 2\sigma_c^2} \right]\,,
\end{equation}
where $L$ is the peak luminosity and $\sigma_c$ the lognormal scatter,
the discrepancies are indeed not that significant. The two methods
both predicted the correct peak luminosities, $L$, of the central
galaxies.  While the lognormal scatters, $\sigma_c$, are slightly
smaller in the intermediate halo mass range and slighter larger in the
lowest halo mass bin at $\sim 0.01$ levels. Note that since the halo
mass estimations in the group catalogs are based on the ranking of
characteristic group luminosity/stellar masses, where the central
galaxy luminosity/stellar mass and halo mass is somewhat correlated
\citep{Yang2008}, the typical uncertainty in the $\sigma_c$
constraints is at $\sim 0.02$ (see their Fig. 4). In addition, in the
CLF/SHAM modelings, the typical $\sigma_c$ assumed in literature also
spans a quite large range, $0.15\sim 0.20$.  Thus, for the general
behaviors of our CLF model predictions for central galaxies, such
amount of differences are expected.  Next for the satellite galaxies,
comparing to the CLF obtained from galaxy groups, our fiducial Match2
method gives very nice CLF model predictions in halos with mass
$\ga 10^{13.5}\msunh$.  While the satellite galaxies for our Match1
method at relatively low mass end are significantly
under-predicted. On the other hand, however, in relatively lower mass
halos, the situation is quite different. Our fiducial Match2 method
over-predicted the CLF at about $\sim$40\% level at
$L\sim 10^{9.5}\Lsunhh$. While the Match1 method prediction is much
better.

Shown in Fig.~\ref{fig:CLF_color} are the CLFs measured separately for
red and blue galaxies, respectively.  Symbols with error bars are
results obtained from SDSS galaxy groups. The lines are results
obtained for our fiducial Match2 method. Even if we did not make
special treatments on the color of galaxies for our neighborhood
abundance matching, we still find very similar color dependence as the
galaxy groups, where massive halos clearly contain more red galaxies
than blue galaxies (both centrals and satellites), while the opposite
applies to low mass halos.  Note that such a halo mass dependence is
indeed quite consistent with the halo quenching mechanism \citep[see
e.g.][and references therein]{Wang2017}.

\begin{figure*}
\center
\vspace{0.5cm}
\includegraphics[height=10.0cm,width=14.0cm,angle=0] {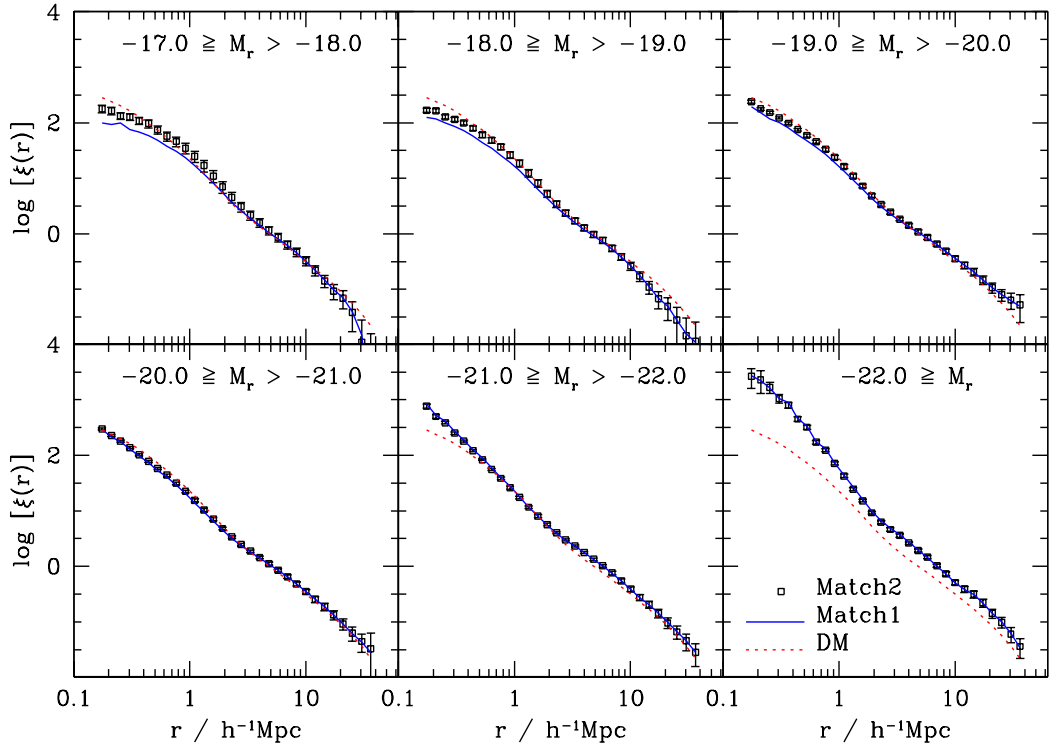}
\caption{Cross correlation functions between galaxies (subhalos) and
  dark matter particles. Different panels correspond to galaxies in
  different absolute magnitude bins as indicated. In each panel, the
  open squares and error bars are results obtained for Match2 method,
  while the solid line are results obtained for Match1 method. For
  reference, the dotted lines indicate the dark matter auto correlation
  in the ELUCID simulation. }
\label{fig:CCF1}
\end{figure*}

\subsection{The conditional stellar mass functions}

Apart from the CLF, which is more observationally related, we can also
measure the CSMF of galaxies.  The CSMF, $\Phi(M_{\ast} \vert M_h)$,
which describes the average number of galaxies as a function of galaxy
stellar mass in a dark matter halo of a given mass, is more
straightforwardly related to theoretical predictions of galaxy
formation models than the CLF, because the conversion from stellar
mass to luminosity in theoretical models requires detailed modeling of
the stellar population and dust extinction.  The CSMF can be estimated
by directly counting the number of galaxies in groups or halos.
However, as the galaxies used here are flux limited, here we need to
take into account the completeness limits of galaxies as a function of
stellar mass as well. 

According to \citet{Bosch2008}, for the stellar masses of the SDSS
galaxies,  at given redshift $z$, the stellar mass is complete above:
\begin{eqnarray} \label{eq:mstarlim}
\lefteqn{\log[M_{*,{\rm lim}}/(h^{-2}\Msun)] =} \\
 & & {4.852 + 2.246 \log D_L(z) + 1.123 \log(1+z) - 1.186 z \over 1 - 0.067
  z} \,. \nonumber
\end{eqnarray}
Using this relation, we can obtain the redshift completeness limit
$z_M$ for a given stellar mass $M_{\ast}$.  Similar to the redshift limit
for luminosities, here we only use galaxies and groups (halos) that
are below redshift limit $z_M$ to estimate the CSMF,
$\Phi(M_{\ast} \vert M_h)$.  In Fig.~\ref{fig:CSMF} we show the
resulting CSMFs for groups of different masses using symbols with
error bars.  The contributions of central and satellite galaxies again
are plotted separately using filled and open symbols.

The solid and dashed lines shown in each panel of Fig.~\ref{fig:CSMF}
are results measured for Match2 and Match1 methods, respectively.
Using the SDSS galaxy group results as references, we find that the
general behaviors of the model predictions of Match2 and Match1
methods are quite similar to those of the CLFs. The model prediction
of Match2 method agrees with that in the SDSS galaxy groups better in
massive halos with mass $\ga 10^{13.5}\msunh$.  While the model
prediction for Match1 method is better in lower mass halos.

In Fig.~\ref{fig:CSMF_color} we show the CSMFs separately for red and
blue galaxies.  Symbols with error bars are results obtained from SDSS
galaxy groups, while the lines are results obtained for our fiducial
Match2 method.  Here again, we see that CSMFs for both red and blue
galaxies can be well recovered.

\begin{figure*}
\center
\vspace{0.5cm}
\includegraphics[height=10.0cm,width=14.0cm,angle=0]{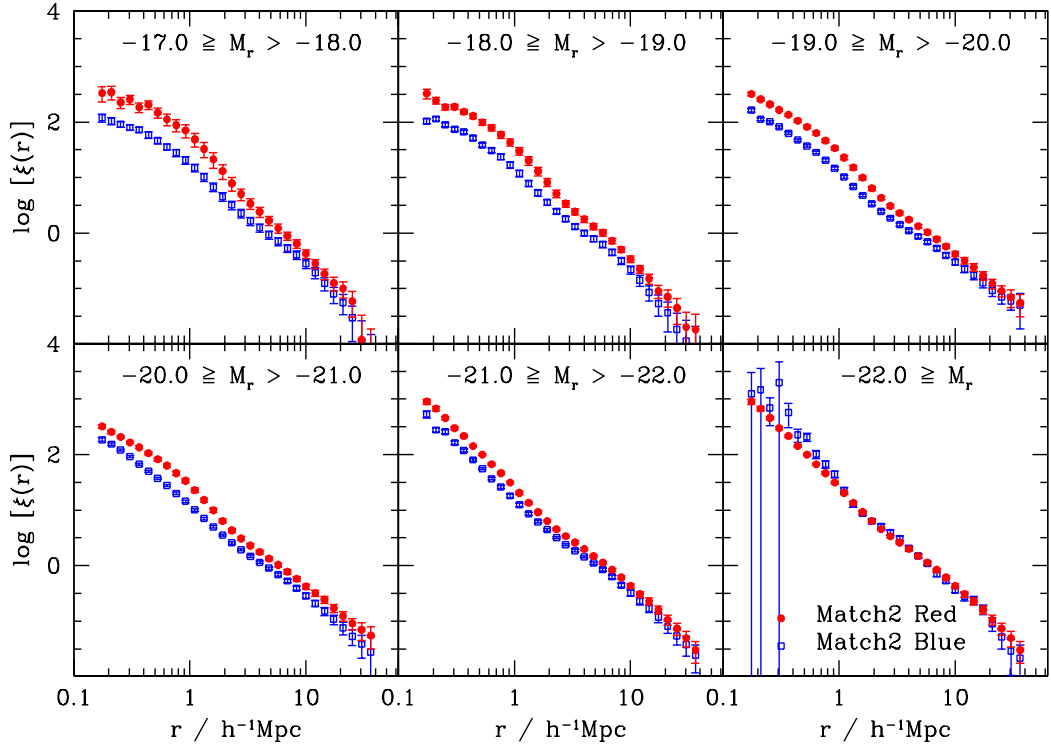}
\caption{Similar to Fig. \ref{fig:CCF1}, but here for galaxies
  (subhalos) that are separated into red and blue subsamples. Here we
  only show results obtained for Match2 method.  }
\label{fig:CCF2}
\end{figure*}

\section{The biases of galaxies}
\label{sec_bias}

Having checked the performance of our galaxy-subhalo connections
established using a neighborhood abundance matching approach in the
HOD framework, we proceed to check their performance on larger scales.
Note that since galaxies are only slightly moved to match nearby main
halos or subhalos, the auto correlation functions of galaxies on large
scales will not change significantly. Here we use the cross
correlation functions between galaxies and dark matter particles to
check if the large scale environment is properly reproduced in the
ELUCID simulation.

\subsection{Cross correlation between galaxies and dark matter}
\label{sec:2PCF}

With all the galaxies been linked with subhalos, we proceed to measure
the cross correlation function (CCF) between subhalos (galaxies) and
dark matter particles,
\begin{equation}%\label{eq:2pcf}
\xi_{\rm CCF} (r) = \frac{P_{\rm HD}(r)}{P_{\rm HR}(r)} -1\,,
\end{equation}
where $P_{\rm HD}(r)$ and $P_{\rm HR}(r)$ are the number of
subhalo-dark matter and subhalo-random pairs, respectively.  For our
investigations, the number of random points has been set to be the
same as the number of dark matter particles within the simulation
box. Those points follow a uniform distribution within the simulation
volume.

We first measure the CCFs between galaxies (subhalos) and dark matter
particles in the ELUCID simulation for overall galaxy population. We
divide the galaxies (subhalos) into 6 subsamples within different
absolute magnitude bins: $-17.0\ge \rmag >-18.0$,
$-18.0\ge \rmag >-19.0$ ...  $-22.0\ge \rmag$.  The open squares shown
in Fig. \ref{fig:CCF1} are the CCFs measured for galaxies within these
absolute magnitude bins for our fiducial Match2 method.  The error
bars are obtained from 100 jackknife re-samplings of the galaxies.  As
a reference, we also show the auto correlation function (ACF) of dark
matter particles in the ELUCID simulation in each panel of
Fig. \ref{fig:CCF1} using a dotted line. Comparing to the ACF of dark
matter, the CCFs of galaxies show somewhat weaker and stronger
clustering strength for fainter and brighter subsamples, which is
qualitatively consistent with those observational measurements of
galaxy biases using ACFs \citep[e.g.][]{Zehavi2005, Wang2007}.
% In general, according to HOD/CLF models \citep[e.g.][]{Yang2003},
% the clustering properties of galaxies at large scales can be
% modelled using a linear bias factor, i.e., the CCFs of galaxies
% should be parallel with the ACF of dark matter particles.  \adb{The
% prominent unparalleled features of CCFs for galaxies with
% $\rmag >-20.0$ at large scales may indicate that these galaxies in
% the SDSS DR7 still suffer significantly from the cosmic variances,
% i.e., because of the existing large void and great wall
% structures. While for brighter galaxies in larger volumes, the
% cosmic variances might be small. }

In addition to the ACF of dark matter particles, for comparison, we
also show using solid line the results obtained for the Match1
method. The main difference induced by the Match2 and Match1 methods
are the satellite fraction of galaxies, especially at the faint
end. According to the CCF comparison, we see that:
\begin{enumerate}
\item The galaxies in the two samples give very similar results on
  large scales at $r\ge 5\mpch$. 
\item On small scales, we see Match2 method has stronger clustering
  strength, especially in the faint galaxy subsamples.
\end{enumerate}
These features indicate that the clustering measurements of galaxies
on small scales are also very important for the HOD/CLF modelling,
especially in constraining the satellite components.

Apart from the CCFs of overall galaxy population, we also measure
the CCFs of galaxies that are separated in to red and blue
subsamples. We show in Fig \ref{fig:CCF2} the CCFs measured separately
for red blue galaxies for the Match2 method using solid dots and open
squares, respectively.  The CCFs of red and blue subsamples show quite
different behaviors where red galaxies show overall stronger
clustering strength than blue galaxies except in the brightest
magnitude bin.  Note that in our neighbor abundance matching approach,
we did not distinguish between red and blue galaxies. Thus the
different clustering behaviors of red and blue galaxies are caused by
their large scale environments, e.g., satellite fraction, and host
halo masses, etc.

Other than the luminosities of galaxies, we also considered galaxies
of different stellar masses. Similar to the treatments for
luminosities, we separate galaxies into 6 subsamples within different
stellar mass bins: $9.0\le \log M_{*} <9.4$, $9.4\le \log M_{*} <9.8$
...  $11.0\le \log M_{*}$.  The clustering properties of galaxies in
different stellar mass bins are very similar to those in different
absolute magnitude bins, which for simplicity, are not shown
explicitly here.

\begin{figure*}
\center
\vspace{0.5cm}
\includegraphics[height=14.0cm,width=14.0cm,angle=0]{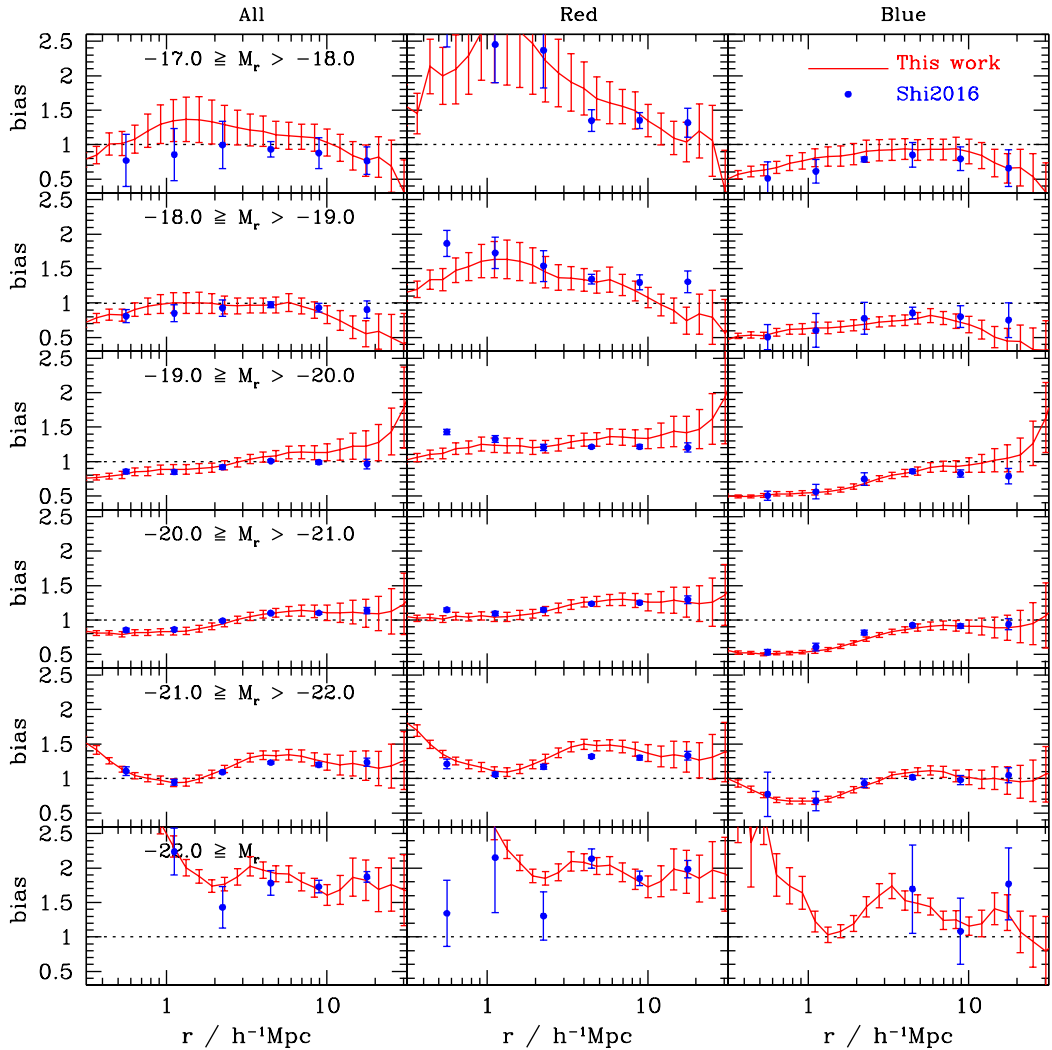}
\caption{The biases for all, red and blue galaxies in different
  luminosity bins as indicted. In each panel, the biases of galaxies
  obtained from the CCFs in this study is shown as the solid line with
  1-$\sigma$ error bars. For comparison, the dotts with error bars are
  results obtained by \citet{Shi2016}. Here again, we only show
  results obtained for Match2 method.  }
\label{fig:bias}
\end{figure*}

To quantify the clustering strengths of galaxies, we show in
Fig. \ref{fig:bias} the ratios of the galaxy-dark matter CCFs and the
dark matter-dark matter ACFs, which are indeed the bias of galaxies as
a function of radius. Here results are shown separately for galaxies
of different colors and in different absolute magnitude bins as
indicated in the plot. The solid line in each panel are results
obtained for our Match2 method. For comparison, we also show the
resulting biases extracted from the reconstructed real space ACFs of
galaxies obtained by \citet{Shi2016} using dots with error bars. In
their study, the redshift space distributions of galaxies are mapped
to real space by correcting redshift distortions on both small and
large scales. Based on thus reconstructed real space distributions of
galaxies, \citet{Shi2016} measured the real space ACFs for galaxies in
different absolute magnitude bins which are the same as used in this
study. The biases of galaxies are then obtained using the square root
of the ratios between the ACFs of galaxies and dark matter particles.

\begin{figure*}
%\plotone{f3.eps}
\center
\vspace{0.5cm}
\includegraphics[height=7.5cm,width=14.5cm,angle=0] {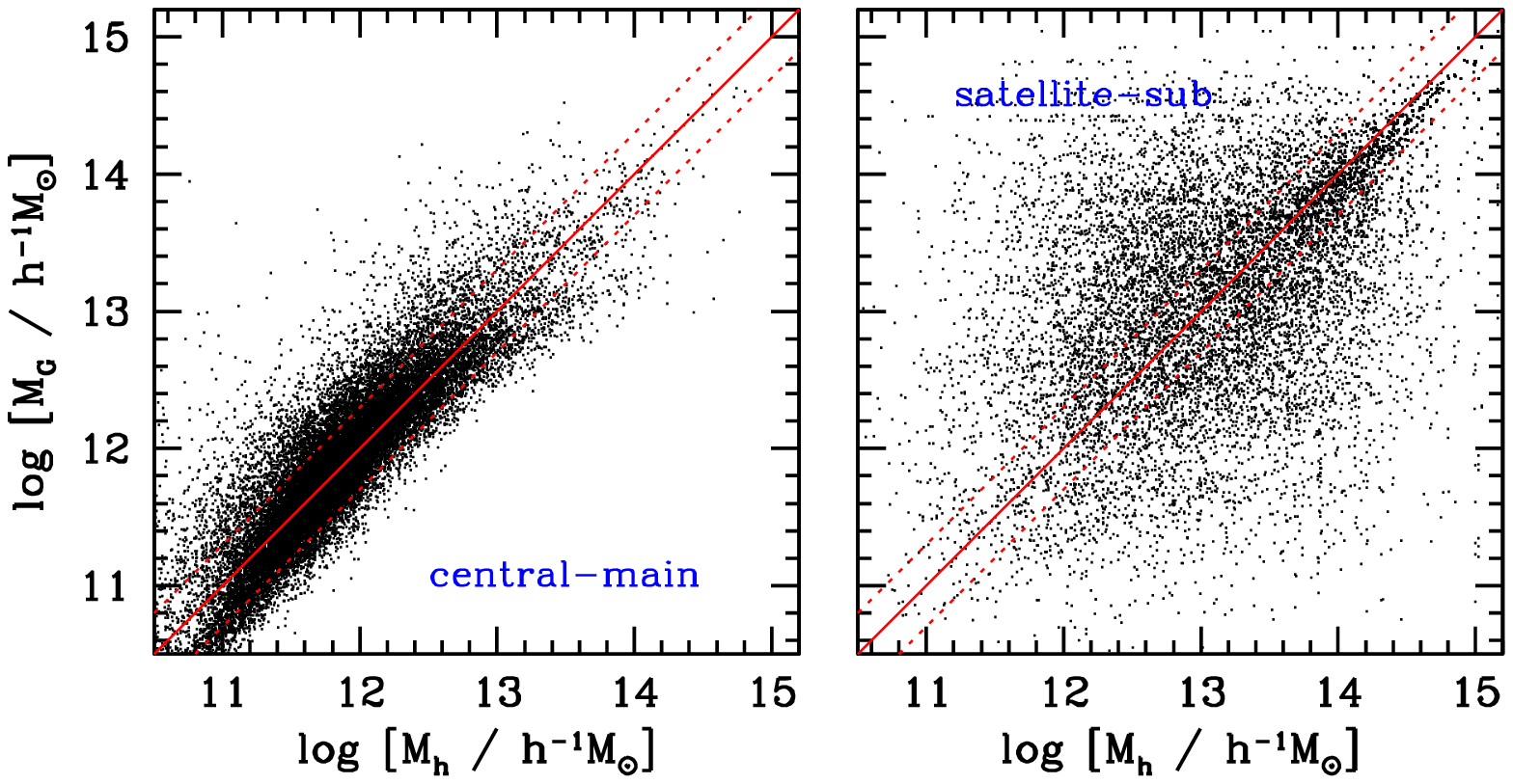}
\caption{The halo mass obtained from galaxy groups $\log M_G$ v.s. the
  host halo mass of the subhalos $\log M_h$ in the ELUCID simulation
  of the matched galaxy-subhalo pairs. Shown in the left and right
  panels are results for the central-main pairs and satellite-subhalo
  pairs, respectively.  }
\label{fig:MM}
\end{figure*}

By Comparing our model predictions with those obtained by
\citet{Shi2016}, we find that these two measurements agree quite well,
especially for all and blue galaxies.  In most cases, the data points
agree with each other within 1-$\sigma$ level, except a few slightly
larger than 1-$\sigma$ level. While the discrepancies are somewhat
larger for red galaxies. There are quite a number of data points that
deviate from each other at about 2-$\sigma$ level.  Apart from these
agreement check, we also find that both of these bias measurements
reveal some curvatures in the $-21.0\ge\rmag >-22.0$ magnitude bin.
According to the error bars, we believe that the curvature around
$1\mpch$ which roughly corresponds to a transition scale from 1-halo
to 2-halo term is robust. The curvature at this scale, which is quite
different for red and blue galaxies, might be useful for galaxy
formation constraints.

The overall agreement of the bias for our galaxy-subhalo matched pairs
indicate again that our neighborhood abundance matching method works
very well and the large scale environments in our ELUCID simulation is
quite reliably reproduced.

\section{How to use the matched data}
\label{sec_use}

Theoretically, if one can provide a perfect link between the observed
galaxies and the subhalos in the simulation, one can then use the
properties of individual galaxies to constrain galaxy formation
models, e.g. via SAMs, etc., to unprecedented precision.  The
galaxy-subhalo connections obtained in this study from the ELUCID
simulation, although not perfect, are already much better than the
traditional subhalo abundance matching approach.

With all the above tests on both small (halo-based) and large scales
for the feasibility and reliability of our neighborhood abundance
matching method, we proceed to provide some suggestions for the use of
the matched galaxy-subhalo connections. Here we suggest to divide all
the galaxy-subhalo pairs into three categories, (1) halo-based pairs,
(2) mass and local volume pairs and (3) local volume pairs.

To make this separation, we first extract all the galaxy-subhalo pairs
that are either central-main pairs or satellite-subhalo pairs. We show
in Fig. \ref{fig:MM} the group mass v.s. halo mass for these two kinds
of galaxy-subhalo pairs.  Shown in the left and right panels are
results for the central-main, satellite-subhalo pairs, respectively.
For the central-main pairs, although we see there are some pairs quite
off from the consistency line which are caused by various reasons,
e.g., survy edge effect, mismatch, etc.  the vast majority are
consistent with each other.  For the satellite-subhalo pairs, the
situation is somewhat worse. We can see that quite a large fraction of
them are quite offset from the consistency line, which are mainly
caused by mismatch of satellite galaxie into different host halos.  As
an illustration, we use two dotted lines $\log M_G-\log M_h=\pm 0.3$
to separate the galaxy-subhalo pairs. In total, there are 212798
central-main and 43178 satellite-subhalo pairs with
$|\log M_G-\log M_h|\le 0.3$.  Comparing to the total number of 277139
central-main and 118930 satellite-subhalo pairs, they consist roughly
77\% and 36\% central and satellite population. If we only consider
galaxies in halos with mass $\log M_h\ge 13.5$, there are about 51\%
and 54\% central and satellite population have
$|\log M_G-\log M_h|\le 0.3$.

According to the above behaviors of galaxy-subhalo pairs, we separate
them into three categories:
\begin{itemize}
\item Cat 1 (halo-based pairs): as pointed out in \citet{Tweed2017},
  the reconstructed simulation can roughly reproduce more than half of
  the halos with mass $\ga 10^{13.5}\msunh$ (e.g., with more than half
  particles in common). Here we select galaxy-subhalo pairs that have
  $\log M_h\ge 13.5$ and $s\le 3\mpch$ (where $s$ is the
  galaxy-subhalo pair separation in redshift space), and
  $\log M_{sh}\ge 11.5$. In total, there are 830 central-main and 7557
  satellite-subhalo pairs fall into this category. In contrast, for
  the same criteria, there are 13 central-main and 108
  satellite-subhalo pairs fall into this category for a rotated
  version of the ELUCID simulation. For these pairs, we suggest that
  one can use the related galaxy properties for those individual main
  or subhalos to evaluate the galaxy properties predicted by SAMs
  individually.
\item Cat 2 (mass and local volume pairs): all the other
  galaxy-subhalo pairs that have $|\log M_G-\log M_h|\le 0.3$.  There
  are 211968 central-main and 38195 satellite-subhalo pairs fall into
  this category. If we only consider galaxies in halos with mass
  $\log M_h\ge 13.5$, the related numbers are 809 for central-main and
  18250 for satellite-subhalo pairs, respectively.  For these pairs,
  we suggest to compare the overall galaxy properties in similar mass
  halos in the same locate volumes, e.g., within radius
  $\sim 20\mpch$.
\item Cat 3 (local volume pairs): all other galaxy-subhalo pairs. For
  these pairs, one may compare the overall galaxy properties
  predicted by SAMs in given spherical regions with radius
  $\sim 20\mpch$ with those SDSS galaxies linked with subhalos in the
  same regions.
\end{itemize}
Based on these criteria, we will evaluate a few SAMs in a subsequent
paper.

\section{Summary}
\label{sec_conclusion}

In this paper, we have proposed a novel neighborhood abundance
matching method to link galaxies in the SDSS DR7 observation with dark
matter main and subhalos in the ELUCID simulation. Here we used two
matching method to make the abundance matching: Match1 is quite
popular in SAMs where galaxies are linked to all the survived main
halo and subhalos, and Match2 where central galaxies are linked with
main halos and satellite galaxies with survived subhalos separately,
all of which the maximum masses of the subhalos along their accretion
histories are used. We made a list of tests on thus established
galaxy-subhalo connections, and the main features are listed a
follows:
\begin{itemize}
\item Based on Match2 method, we measured and modelled the luminosity
  (stellar mass) - subhalo mass relations for central and satellite
  galaxies separately and found that they have quite different
  behaviors.
\item  We have checked the satellite fractions of galaxies as a
    function of luminosity and stellar mass and found Match1 method
    somewhat underestimates the related values, especially for low
    mass galaxies. In addition, unlike the Match2 method, the color
    segregation of satellite fraction is not well reproduced in Match1
    method. 
  \item We have measured the CLFs and CSMFs of galaxies in halos of
    different masses.  Compare to the observational results, the model
    prediction of Match2 method agrees with that in the SDSS galaxy
    groups better in massive halos with mass $\ga 10^{13.5}\msunh$.
    While the model prediction for Match1 method is better in lower
    mass halos.
\item We have measured the biases of galaxies as a function of radius,
  which show overall quite nice agreement with the observational
  results obtained by \citet{Shi2016}. 
\item We have also checked the above quantities separately for red and
  blue galaxies. All of the results for our Match2 method agree with
  the direct measurements from observation fairly well.
\end{itemize}

The above tests show that the Match2 method performs somewhat better
than the Match1 method.  We thus suggest to make use of the
galaxy-subhalo connections established in this sample for galaxy
formation studies, e.g. SAMs performed on the ELUCID simulation.  In
addition, we suggest that those galaxy-subhalo pairs can be divided
into three categories: (1) halo-based pairs which can be used to
evaluate galaxy properties in individual subhalos, (2) mass and local
volume pairs which can be used to evaluate the overall galaxy
properties in similar mass halos in the same local volumes, and (3)
local volume pairs which can be used to evaluate the overall galaxy
properties in the same small volumes.  Finally, the galaxy-subhalo
links and the subhalo merger trees in the SDSS DR7 region for our
ELUCID simulation are available upon request.

%%%%%%%%%%%%%%%%%
% Ackowledgements
%%%%%%%%%%%%%%%%%

\section*{Acknowledgments}

We thank the anonymous referee for helpful comments that greatly
improved the presentation of this paper.  This work is supported by
the 973 Program (No. 2015CB857002), national science foundation of
China (grant Nos. 11233005, 11421303, 11522324, 11503064, 11621303,
11733004) and Shanghai Natural Science Foundation, Grant No.
15ZR1446700. We also thank the support of the Key Laboratory for
Particle Physics, Astrophysics and Cosmology, Ministry of Education.
HJM would like to acknowledge the support of NSFC-11673065 and NSF
AST-1517528, and FvdB is supported by the US National Science
Foundation through grant AST 1516962. WC is supported by the {\it
  Ministerio de Econom\'ia y Competitividad} and the {\it Fondo
  Europeo de Desarrollo Regional} (MINECO/FEDER, UE) in Spain through
grant AYA2015-63810-P as well as the Consolider-Ingenio 2010 Programme
of the {\it Spanish Ministerio de Ciencia e Innovaci\'on} (MICINN)
under grant MultiDark CSD2009-00064.

A computing facility award on the PI cluster at Shanghai Jiao Tong
University is acknowledged. This work is also supported by the High
Performance Computing Resource in the Core Facility for Advanced
Research Computing at Shanghai Astronomical Observatory.

%%%%%%%%%%%%%%%
% Bibliography
%%%%%%%%%%%%%%%

\label{lastpage}

\end{document}